\documentclass[final,3p,times]{elsarticle}
\pdfoutput=1
\usepackage{amsmath}
\usepackage{amssymb}

\usepackage{draftwatermark}
\SetWatermarkScale{50}

\usepackage{caption}
\usepackage{subcaption}
\newcommand{\abs}[1]{\left\vert#1\right\vert}

\newcommand{\brac}[1]{\left( #1 \right)}
\newcommand{\sqbrac}[1]{\left[ #1 \right]}
\newcommand{\pde}[2]{\frac{\partial #1}{\partial #2}}

\usepackage{tikz}

\usepackage{mathptmx}

\journal{XXX}

\begin{document}

\begin{frontmatter}


  \title{Multi-physics simulations of lightning strike on
    elastoplastic substrates.}

\author{Dr S.\ Millmore and Dr N.\ Nikiforakis}

\address{Cavendish Laboratory, Department of Physics, University of Cambridge}

\begin{abstract}
  This work is concerned with the numerical simulation of
  elastoplastic, electromagnetic and thermal response of aerospace
  materials due to their interaction with a plasma arc under lightning
  strike conditions.  Current approaches treat the interaction between
  these two states of matter either in a decoupled manner or through
  one-way coupled `co-simulation'.  In this paper a methodology for
  multiphysics simulations of two-way interaction between lightning
  and elastoplastic materials is presented, which can inherently
  capture the non-linear feedback between these two states of matter.
  This is achieved by simultaneously solving the magnetohydrodynamic
  and the elastoplastic systems of equations on the same computational
  mesh, evolving the magnetic and electric fields dynamically.  The
  resulting model allows for the topological evolution and movement of
  the arc attachment point coupled to the structural response and
  Joule heating of the substrate.  The dynamic communication between
  the elastoplastic material and the plasma is facilitated by means of
  Riemann problem-based ghost fluid methods.  This two-way coupling,
  to the best of the authors' knowledge, has not been previously
  demonstrated.  The proposed model is first validated against
  experimental laboratory studies, demonstrating that the growth of
  the plasma arc can be accurately reproduced, dependent on the
  electrical conductivity of the substrate.  It is then subsequently
  evaluated in a setting where the dynamically-evolved properties
  within the substrate feed back into the plasma arc attachment.
  Results are presented for multi-layered substrates of different
  materials, and for a substrate with temperature-dependent electrical
  conductivity. It is demonstrated that these conditions generate
  distinct behaviour due to the interaction between the plasma arc and
  the substrate.
\end{abstract}


\end{frontmatter}

\section{Introduction}
\label{sec:introduction}

On average, every commercial airliner is struck by lightning once a
year, hence all aircraft undergo rigorous testing to ensure these
strikes do not lead to major in-flight
damage~\cite{morgan2012interaction}.  Traditionally, most aircraft
skins have been made from aluminium, which is lightweight and strong
under normal in-flight conditions, but also both thermally and
electrically conductive, thus it quickly dissipates the energy
deposition from a lightning strike away from the impact site.  Modern
designs increasingly make use of carbon composite materials, which are
stronger and lighter than aluminium under normal aircraft operating
conditions~\cite{segui2014boeing}.  However, they have much lower
thermal and electrical conductivity, which leads to increased energy
deposition at the site of a lightning strike, due to the Joule heating
effect, which in turn can lead to much greater damage on an aircraft
panel.  In order to mitigate these effects, composite materials
typically include an interwoven wire fabric, threads of high
conductivity wires which dissipates current away from the initial
impact site, reducing the local energy deposition.  However, this
increases the weight of the aircraft, negating some of the savings
introduced through the lightweight
composite.

\begin{figure}[!ht]
  \centering
  \includegraphics[width=0.4\textwidth]{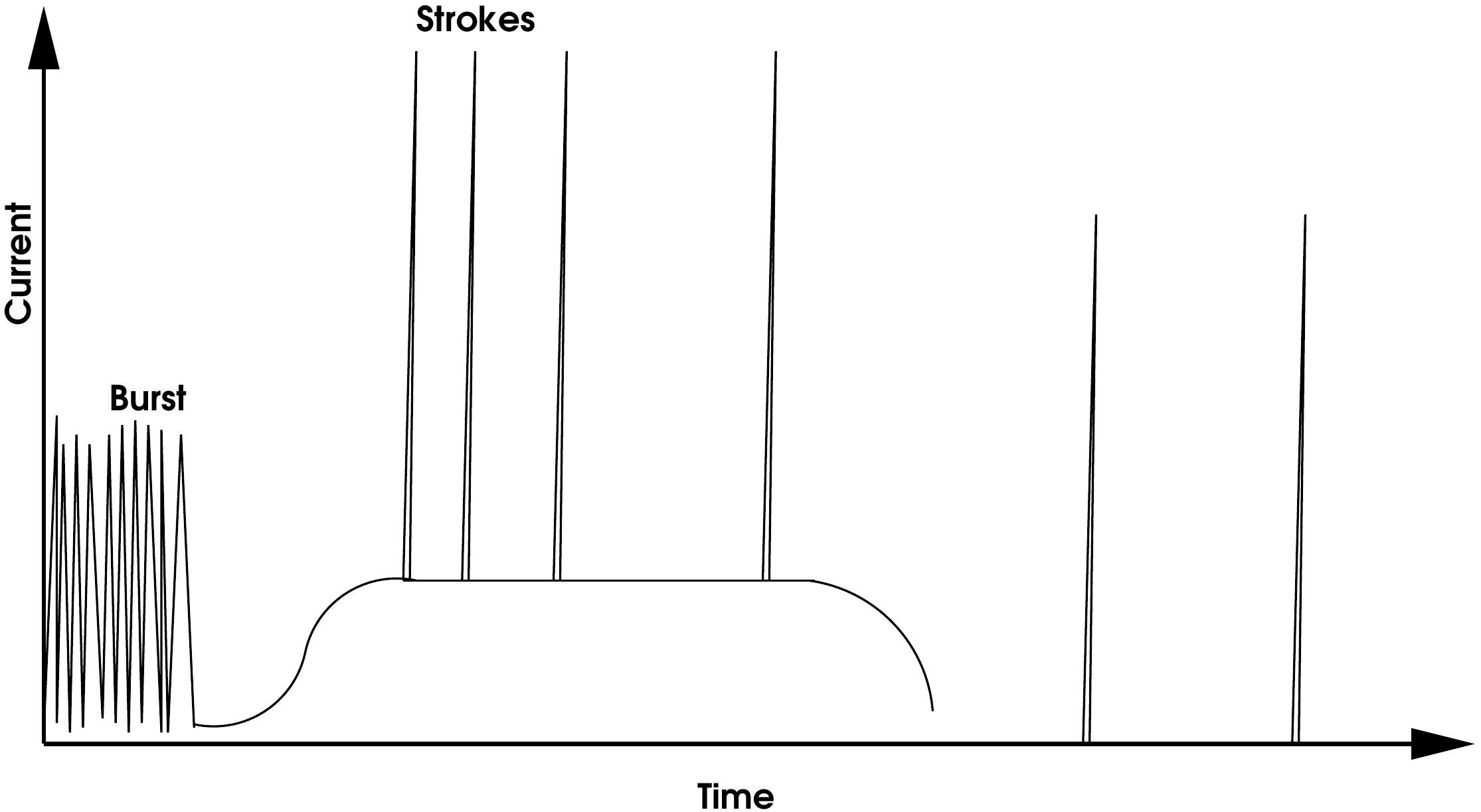}
  \caption{Illustration of the typical features in the current over
    the course of a Lightning strike, adapted
    from~\cite{lalande1999analysis}.  There is an initial burst, as
    the stroke attaches (about 3\,ms duration), followed by a longer
    period of continuous current input (around 200\,ms and 330\,A).
    Superimposed upon this longer input are high-current strokes, with
    a period of around 20\,ms and currents which can exceed
    200,000\,A.  These strokes can continue after the continuous
    current input ends.}
  \label{fig:strike-profile}
\end{figure}

Over the course of a lightning strike, the current flow consists of a
long continuing current of a few hundred amperes, which can last for
up to 1\,s.  Superimposed upon this are multiple high-current peaks,
each with duration less than 0.5\,ms, these are referred to as
`strokes', illustrated in Figure~\ref{fig:strike-profile}.  This
information is used to create standardised wave forms for experimental
studies~\cite{ARP5412B} which are designed to be representative of
severe conditions and each stroke can reach maximum current input
between 100,000\,A and 200,000\,A.  Such extreme conditions are
expected for less than 5\% of lightning strikes.  Damage due to
lightning strike falls into two categories, direct and indirect
effects.  Direct effects are localised damage due to the arc
connection, for which the individual strokes, shown in
Figure~\ref{fig:strike-profile}, are the primary cause.  Indirect
effects involve the electrodynamic interaction of the process with the
entire aircraft, and therefore consider the entire strike
profile~\cite{arp20045416}.

In this work, the local damage to the substrate, caused by an
individual stroke, is the primary area of interest.  Experimental
modelling of these effects typically uses small-scale
investigations~\cite{arp20045416} over the duration of a single
stroke.  Though at a much smaller scale than a full lightning strike,
by using extreme stroke conditions, these provide a good model of
lightning strike damage.  The advantage of these experiments is that
the impact point on the substrate can be closely controlled, allowing
analysis over the entire course of the experiment.  Numerical
simulations can reduce the expense of repeated experiments and can
complement experimental measurements during the dynamic interaction
between a lightning strike and a substrate.

The interaction between the arc and the substrate is a complex,
non-linear process, and thus presents challenges in capturing the full
behaviour within a numerical model.  Initial work in simulating the
arc profile was developed from a numerical magnetohydrodynamic (MHD)
description of an argon arc by Hsu {\em et al}.~\cite{hsu1983study},
with applications in plasma arc welding.  Gliezes {\em et al}.\ also
considered stationary arcs, and this allowed a temperature profile at
the attachment point to be computed~\cite{gleizes2005thermal}.  The
turbulent motion of the arc channel was simulated by Chemartin {\em et
  al}.~\cite{chemartin2009three}, and a smaller scale arc, with
current profile given between an anode and a cathode (the substrate
skin) was later developed by Chemartin {\em et
  al}.~\cite{chemartin2011modelling}.  This model was then applied to
a swept arc, with multiple attachment
points~\cite{chemartin2012direct}, and a subsequent analysis of the
effects this had on a material substrate was carried out.

Villa {\em et al}.~\cite{villa2011multiscale} consider the pressure
loading above a substrate using an MHD plasma model, with a prescribed
current density evolution.  Although pressures above the substrate are
measured, the effect within the substrate itself was not
investigated.  The effects of the conductivity of the cathode were
considered by Tholin {\em et al}.~\cite{tholin2015numerical}, and this
showed how the shape of the arc attachment changes dramatically for
low conductivity carbon composite materials.  This work was expanded
upon in the thesis of Martins~\cite{martins2016etude}, which focuses
on experimental measurements of plasma arcs using a variety of
substrates.

Typically, the effects of the arc attachment on the substrate are
modelled separately. Ogasawara {\em et al}.~\cite{OGASAWARA2010973},
consider the damage within a carbon composite substrate, though do not
model the plasma arc directly, but instead use a prescribed current
input.  Abdelal and Murphy~\cite{ABDELAL2014268} also take this
approach, with modifications as to how the current profile is applied,
as did Guo {\em et al}.~\cite{GUO201710}.  Foster {\em et
  al}.~\cite{FOSTER2018364} highlight the importance of movement of
the attachment point through prescribed motion of the current profile.
Karch {\em et
  al}.~\cite{iet:/content/conferences/10.1049/ic.2015.0149} compute
damage to a carbon composite substrate through a prescribed expansion
of a plasma arc expansion.

The computational models described above typically simulate the arc
attachment process and the substrate response individually, rather
than as a two-way interacting system, and coupling is achieved through
a `co-simulation' approach, modelling each system individually.  The
Joule heating and pressure loading effects of the arc attachment lead
to damage within the substrate.  However, this can both alter the
shape of the substrate (either through bending or damage) and can
change the properties of the substrate, such as electrical
conductivity, which leads to a feed-back effect changing the arc
attachment.  Therefore a truly nonlinear multi-physics approach is
needed to capture the two-way interaction between the two systems.
The approaches described above do not capture this behaviour in a
single model, and hence cannot fully replicate these non-linear
effects.

In this work, a multi-physics methodology is presented which allows
for the dynamic non-linear coupling of the plasma arc and the
substrate.  The framework developed within the Laboratory for
Scientific Computing at the University of Cambridge is
used~\cite{Schoch2013163,MICHAEL20181}, which simultaneously solves
coupled elastoplastic and fluid equations.  This framework is extended
to simulate the interaction between a MHD description of a plasma arc
and the elastoplastic equations.  Through this, the feedback between
the two states of matter can be captured; the plasma arc alters the
properties of the substrate, and this in turn affects the topology of
the arc.

The rest
of the paper is laid out as follows: In Section~\ref{sec:math-form}
the mathematical formulation of the model components, and the
multi-material coupling, is detailed.  In
Section~\ref{sec:validation}, validation of the plasma model used in
this work is presented.  In Section~\ref{sec:multi-layer-substr} the
coupling to multi-layered substrates is demonstrated, and in
Section~\ref{sec:temp-depend-cond} the ability for this model can
capture feedback from the substrate into the plasma arc is shown.
Conclusions and further work are given in
Section~\ref{sec:conclusions}.

\section{Mathematical formulation}
\label{sec:math-form}

\begin{figure}[!ht]
  \centering
  \includegraphics[width=0.48\textwidth]{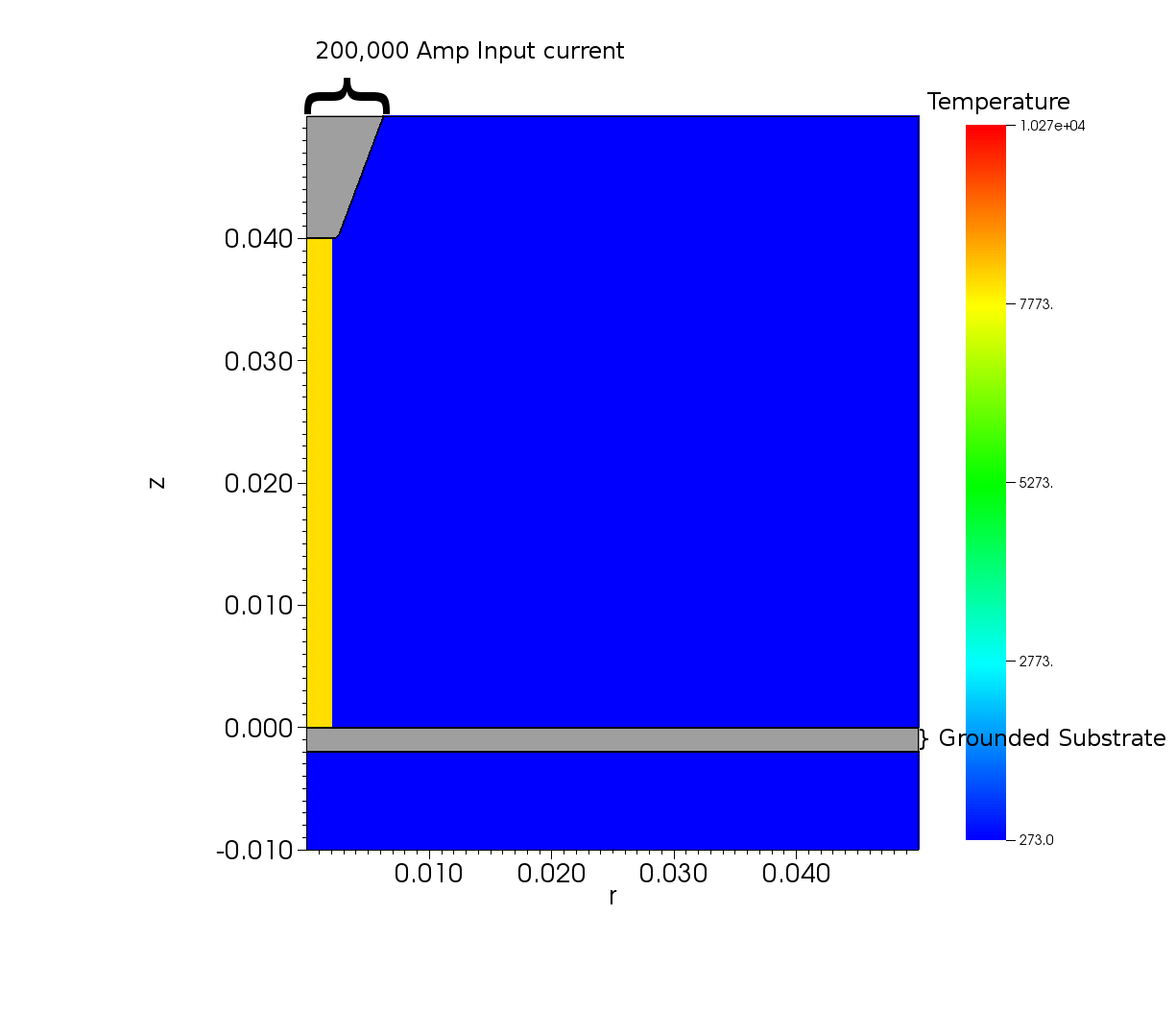}
  \caption{Schematic of an air plasma arc generated by an electrode
    connecting with a grounded conductive substrate. }
  \label{fig:arc-substrate-example}
\end{figure}

In this section, the mathematical models used to describe the
interaction between a plasma arc and an elastoplastic substrate are
presented.  A reference configuration to describe the application of
this model is considered; a plasma arc in air, generated by an
electrode, and connected to a conductive elastoplastic substrate which
is grounded at its outer edges.  This is representative of the
laboratory framework for testing the effects of lightning strike on
aircraft skin configurations, such as those used
in~\cite{villa2011multiscale,tholin2015numerical,martins2016etude}.
Within this framework, a cylindrically symmetric model is considered,
with the arc connection at the centre of the domain, which is
sufficient to capture the bulk behaviour of the arc-substrate
interaction~\cite{martins2016etude}. This configuration is illustrated
in Figure~\ref{fig:arc-substrate-example}, for which a single-material
isotropic substrate is shown.  A blunt electrode is placed above a
grounded substrate, and sufficient current is passed through the
electrode to generate a plasma arc.  The voltage breakdown of the air,
which occurs at timescales much shorter than the mechanical evolution
of the system, is not modelled within this framework.  Instead the
procedure of e.g.\ Chemartin {\em et
  al}.~\cite{chemartin2011modelling}, Larsson {\em et
  al}.~\cite{larsson2000lightning}, and Tholin {\em et
  al}.~\cite{tholin2015numerical}, is followed, and a thin pre-heated
region of the domain is considered, representative of the initial
connection resulting from voltage breakdown.  This region is of
sufficiently high temperature to result in ionisation and the
formation of a plasma.  This allows current flow from the electrode to
the grounded edge of the substrate.  It has been shown that the values
within this preheated region do not affect the overall evolution of
the plasma arc~\cite{chemartin2011modelling,larsson2000lightning}.

When considering the MHD approach to modelling plasma, it is generally
assumed that the arc is close to local thermodynamic equilibrium
(LTE), i.e.\ the arc can be described with a single temperature.  This
has been shown to give good agreement with experimental
studies~\cite{tholin2015numerical,martins2016etude}, and the
complexity in otherwise determining transport coefficients for a
non-equilibrium air plasma means the LTE approach is widely used, thus
this approach is used in this work. Simulations using non-equilibrium
plasmas do exist~\cite{0963-0252-14-1-016}, but the added complexity
of these models, in which detailed chemistry is required for each
species modelled, means they are not generally used when the LTE
approach holds.

The substrate and electrode materials are described through an
elastoplastic model in the Eulerian
frame~\cite{Schoch2013163,MICHAEL20181}.  The domain beneath the
substrate is included such that it is straightforward to incorporate
the grounding of the substrate only at the outer edge.  This domain is
considered to be air, described through an ideal gas law, since over
the timescales considered, it is not heated sufficiently for
ionisation and a plasma model to be required.

\subsection{The plasma model}

Under the LTE assumption, the plasma arc is described through the
single fluid Euler equations coupled with the Maxwell equations for
electrodynamic fields.  This gives three equations for the
conservation of mass, momentum and energy,

\begin{subequations}
  \label{eq:villa-model-eqs}
  \begin{align}
    \label{villa-mhd-rho}
    \pde{\rho}{t} + \pde{}{x_i} \brac{\rho u_i} &= 0 \\
    \label{villa-mhd-mom}
    \pde{}{t}\brac{\rho u_j} + \pde{}{x_i} \brac{\rho u_i u_j +
      \delta_{ij}p}  &= \brac{{\bf J} \times {\bf B}}_j \\
    \pde{U}{t} + \pde{}{x_i} \sqbrac{u_i\brac{U + p}} &= u_i \brac{{\bf J} \times {\bf B}}_i + \eta J_i J_i - S_r , 
    \label{villa-mhd-e}
 \end{align}
\end{subequations}
where $\rho$ denotes the density, ${\bf u}$ the velocity vector, $p$
the pressure and $U$ the total energy.  The source term
${\bf J} \times {\bf B}$ is the Lorentz force due to circulation of
the electric current, where ${\bf J}$ is the current density and
${\bf B}$ the magnetic field.  The source term
$\eta {\bf J} \cdot {\bf J} = {\bf J} \cdot {\bf E}$ is the Joule
heating term due to circulation of current in resistive media, where
$\eta = 1 / \sigma$ is the resistivity, the inverse of the electrical
conductivity, and ${\bf E}$ is the electric field.  The final source
term, $S_r$ is a radiative transfer term and in this work, a grey body
treatment is used, following Villa {\em et
  al}.~\cite{villa2011multiscale}.  This is a simplified, temperature
dependent radiative model, though it may not be suitable for
sufficiently large temperature variations.  Improvements to the
radiative model will be the subject of future work.

The electrodynamic source terms are calculated under the assumption
that the electric field is static, depending only on the charge
distribution and voltage gradient.  The conservation of current
density can therefore be written as
\begin{equation}
  \label{eq:cons-j}
  - \nabla \cdot {\bf J} = - \nabla \cdot \brac{\sigma {\bf E}} =
  \nabla \cdot \brac {\sigma \nabla \phi} = 0
\end{equation}
where $\phi$ is the electric potential.  Note that although there is
no explicit time dependence to the electric field, the electrical
conductivity of the plasma is dependent on temperature (and pressure),
thus does have implicit dependence.  The magnetic field is computed
from the current density through the Maxwell-Ampere equation
\begin{equation}
  \label{eq:max-amp}
  {\bf B} = \nabla \times {\bf A}, \qquad \nabla \cdot \nabla A_i =
  -\mu_0 J_i
\end{equation}
where ${\bf A}$ is the magnetic vector potential.

In order to close the equations, an equation of state based on the
work of d'Angola {\em et al}.~\cite{d2008thermodynamic} is used.  This
describes the composition of an air plasma considering the 19 most
important components over temperatures $T < 60,000$\,K and pressures
$0.01 < p < 100$\,atm.  From this, the thermodynamic and
electrodynamic properties of the plasma are also given.  These
relationships are given as fitted functions of pressure and
temperature, which are not invertible.  Therefore for numerical
purposes, this data has been tabulated providing an efficient means to
convert between variables within the current model~\cite{frederik}.  

\subsection{The elastoplastic model}
\label{sec:elastoplastic-model}

The elastoplastic substrate and electrode are described using the
Eulerian framework as presented by Schoch {\em et
  al}.~\cite{Schoch2013163} and Michael {\em et
  al}.~\cite{MICHAEL20181}, based on the formulation of Godunov and
Romenskii~\cite{godunov1972nonstationary}.  Plasticity effects are
incorporated following the work of Miller and
Collela~\cite{miller2001high}.

Since an Eulerian framework is used, the deformation of the solid
materials cannot be described through mesh distortion.  Instead this
behaviour is accounted for through consideration of the deformation
gradient tensor, given by
\begin{equation}
  \label{eq:elas-def-grad}
  \mathrm{F}_{ij} = \frac{\partial x_i}{\partial X_j}.
\end{equation}
This allows mapping back to the original configuration, with
coordinates given by ${\bf X}$ to the deformed configuration,
${\bf x}$.  The technique of Rice~\cite{RICE1971433} is followed, in
which the plastic deformation is considered separately, ${\bf F}^p$,
which means the total deformation can be decomposed into plastic and
elastic components, ${\bf F} = {\bf F}^e{\bf F}^p$.  The evolution of
the solid materials is described by a hyperbolic system of
conservation laws,
\begin{equation}
  \frac{\partial \rho u_i}{\partial t} + \frac{\partial}{\partial
    x_k}\left(\rho u_iu_k - \sigma_{ik}\right) = 0
  \label{eq:solid-form-1}
\end{equation}
\begin{equation}
  \frac{\partial \rho E}{\partial t} + \frac{\partial}{\partial
    x_k}\left(\rho E u_k - u_i\sigma_{ik}\right) = \frac{1}{\eta} J_i J_i
  \label{eq:solid-form-2}
\end{equation}
\begin{equation}
  \frac{\partial \rho \mathrm{F}_{ij}^e}{\partial t} +
  \frac{\partial}{\partial x_k}\left(\rho u_k\mathrm{F}_{ij}^e - \rho
    u_i\mathrm{F}_{kj}^e\right)
  = -u_i\frac{\partial \rho \mathrm{F}_{kj}}{\partial x_k} + \mathrm{P}_{ij}
  \label{eq:solid-form-2_5}
\end{equation}
\begin{equation}
  \label{eq:solid-form-3}
  \frac{\partial \rho \kappa}{\partial t} + \frac{\partial}{\partial
    x_i} \left( \rho u_i \kappa \right) = \rho \dot{\kappa}
\end{equation}
where $\sigma$ is the stress tensor and $\kappa$ is the scalar
material history parameter which tracks work hardening of the material
through plastic deformation.  The density is related to the
deformation gradient through
\begin{equation}
  \label{eq:solid-density}
  \rho = \frac{\rho_0}{\mathrm{det}\,{\mathrm{\bf F}}^e}
\end{equation}
and the stress tensor is given by
\begin{equation}
  \label{eq:stress-defn}
  \sigma_{ij} = \rho \mathrm{F}^e_{ik} \frac{\partial e}{\partial \mathrm{F}^e_{jk}}
\end{equation}
where $e$ is the specific internal energy.  In order to close the
system, an analytic constitutive model relates the specific internal
energy to the deformation gradient, entropy and material history
parameter, i.e.\ $e = e\brac{{\bf F}^e, S, \kappa}$.

The effects of the current density passing through the solid substrate
can be modelled through a Joule heating term in the energy
conservation law~(\ref{eq:solid-form-2}).  As with the plasma, the
electric field is assumed static in the substrate, and the relevant
equations~(\ref{eq:cons-j}) and (\ref{eq:max-amp}) apply here too.

The system of evolution
equations~(\ref{eq:solid-form-1})--(\ref{eq:solid-form-3}) is coupled
with compatibility constraints, which ensure that deformations remain
physical and continuous, given by
\begin{equation}
  \label{eq:compat-const}
  \frac{\partial \rho \mathrm{F}_{ij}}{\partial x_j} = 0.
\end{equation}

The MHD and elastoplastic solid formulations described in this section
are solved numerically using high-resolution shock-capturing methods
as described in previous work~\cite{Schoch2013163,MICHAEL20181}.

\subsection{The multimaterial approach}
\label{sec:mult-appr}

In this work ghost fluid methods are used, in combination with level
set methods, to model the interfaces between the plasma arc, or air,
and the substrate and electrode.  Level set methods track the
evolution of the interfaces, as they evolve over time, e.g.\ substrate
bending under the impact loading of the plasma arc.  In order to
provide boundary conditions at these interfaces, the Riemann ghost
fluid method is used, which solves mixed material Riemann problems to
give interface states during evolution of the governing equations.

Level set methods represent the interface between a pair of materials
as a signed distance function, $\phi\brac{\bf x}$, with the zero
contour of this function being the physical location of that
interface.  It is assumed that there is no mass transfer between
materials, and this gives an advective law for evolving the level set
function,
\begin{equation}
  \label{eq:ls-evol}
  \frac{\partial \phi}{\partial t} + {\bf u} \cdot \nabla \phi = 0
\end{equation}
where ${\bf u}$ is the material velocity.  This equation is evolved
using a third order Hamilton-Jacobi WENO reconstruction
scheme~\cite{levelsets}.  Under a non-uniform velocity field, the
level set function will not remain a signed distance function without
reinitialisation.  Each material within the model is assigned a level
set function, and a fast marching algorithm to preserve the signed
distance function around the contour $\phi\brac{\bf x} = 0$ is used.
The physical material for a given point can then be determined through
identifying the single positive level set function.

The Riemann ghost fluid method, developed by Sambasivan and
Udaykumar,~\cite{2009AIAAJ..47.2907S}, provides dynamic boundary
conditions at the material interfaces, based on the original method of
Fedkiw {\em et al}.~\cite{Fedkiw}.  To provide these conditions,
following procedure is used for each material, $m$,

\begin{enumerate}
\item For a cell $i$, if $\phi_{i,m} < 0$ and an adjacent cell
  $\phi_{i\pm,m} > 0$, it is adjacent to the interface, the closest
  interfacial location is given by $P = i - \phi_m {\bf n}$
\item Two probes are projected into the two adjacent materials, to the
  points $P_L = P + {\bf n} \cdot \Delta{\bf x}$ and  $P_R = P - {\bf
    n} \cdot \Delta{\bf x}$
\item States ${\bf W}_{L}$ and ${\bf W}_{R}$ are interpolated for each
  of these points from the surrounding cells
\item A mixed material Riemann problem is solved to obtain the star
  state ${\bf W}_L^*$
\item The material state in cell $i$ is replaced by ${\bf W}_L^*$
\end{enumerate}

In this procedure, it is assumed, without loss of generality, that the
state ${\bf W}_L$ is the material that exists for $\phi_m > 0$.  It is
noted that once the states $P_L$ and $P_R$ are found vector
quantities, ${\bf W}_L$ and ${\bf W}_R$ must be projected into
components normal and tangential to the interface.  Once all
interfacial cells have been assigned a boundary value, a fast marching
method is used to fill the region $\phi_m$ such that the stencil of
the numerical method is always satisfied.  The mixed-material Riemann
problems are based on linearised solutions to the systems of
equations, and are described in~\cite{MICHAEL20181}.

The electrodynamic quantities, current density and magnetic field, are
assumed to be continuous across material boundaries, and thermal
effects due to temperature differences in the substrate materials and
the plasma arc are not modelled.  The governing equations
(\ref{eq:cons-j}) and (\ref{eq:max-amp}) are solved across the entire
domain for all materials, rather than on a per-material basis.

\section{Validation}
\label{sec:validation}

Experimental validation studies for lightning strike plasma arc
interaction face difficulties in capturing the arc development due to
the likelihood of damage to electronic equipment from the strong
current which generates the arc.  By recording features of the arc at
a sufficient distance can overcome these difficulties, for example,
Martins~\cite{martins2016etude} uses light emission from the arc,
whilst Villa, Malgesini and Barbieri~\cite{villa2011multiscale} use
pressure gauges away from the arc.  We can use these results to
validate the current approach, and henceforth these two models shall
be referred to as M16 and VMB11 respectively.

\subsection{Validation of the MHD equations}
\label{sec:valid-against-press}

\begin{figure}[!ht]
  \centering
  \includegraphics[width=0.4\textwidth]{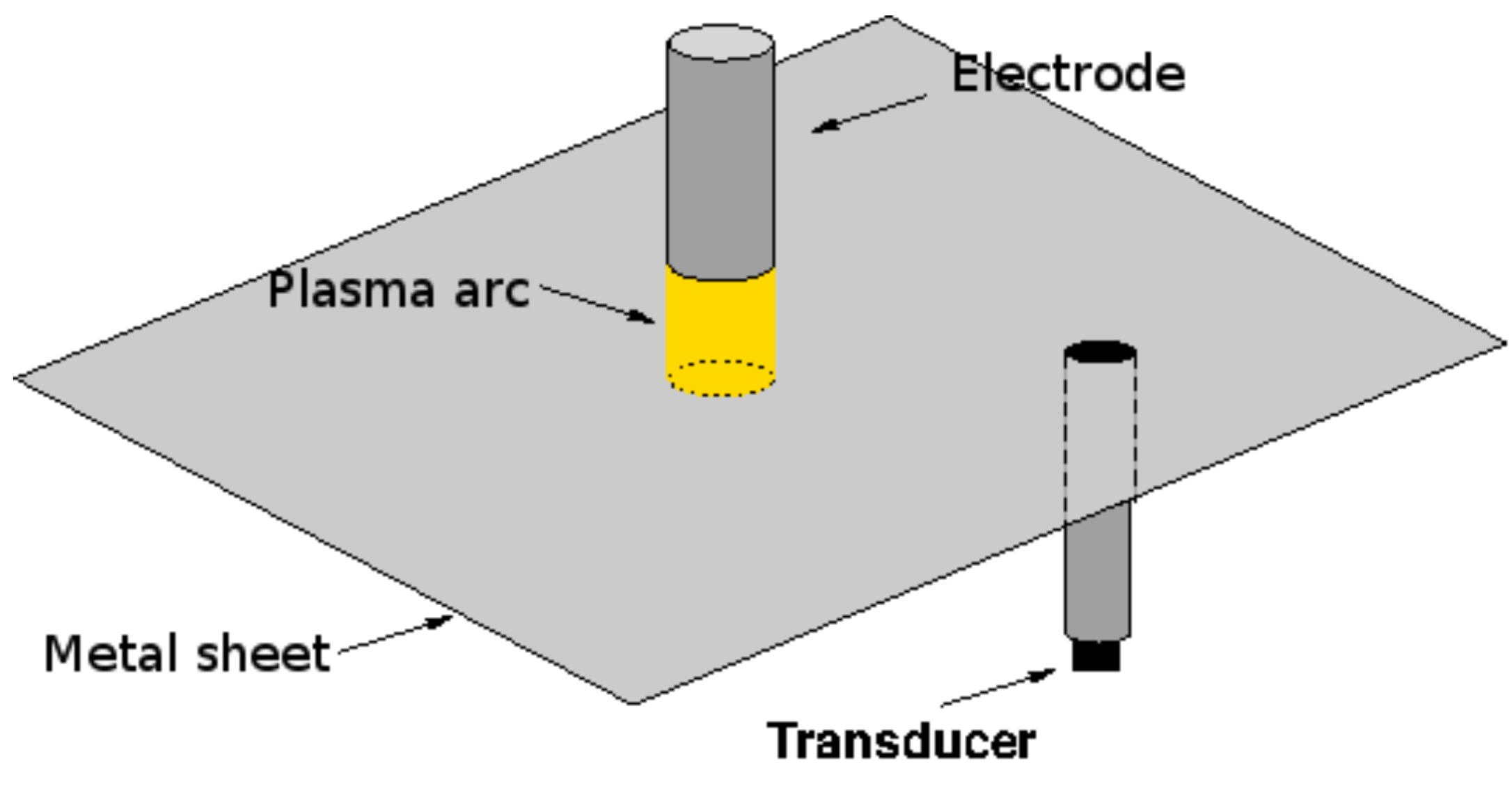}
  \caption{Experimental set-up as used by Villa {\em et
      al}.\cite{villa2011multiscale}.  A plasma arc is generated
    between an electrode and a metal plate.  Away from the attachment
    point, thin tubes are mounted, with pressure sensors at the end of
    these tubes.}
  \label{fig:villa-expt}
\end{figure}

The work in VMB11 allows the MHD equations~(\ref{eq:villa-model-eqs})
to be validated under a given current density profile.  In
Figure~\ref{fig:villa-expt}, the VMB11 configuration is shown, in
which a cylindrical electrode is used to generate a plasma arc, which
connects to a flat metal sheet grounded at its outer edges.  Tubes are
mounted to this sheet in three locations, and a pressure-recording
transducer is placed at the end of these tubes.  As the expanding
plasma arc travels over these tubes, it generates a pressure wave
which travels down the tubes, and is recorded by the transducer.

To simulate this experiment, cylindrical symmetry is used, and the
electrode is placed 5\,cm above a reflective boundary.  The current
flow through from the electrode was recorded to follow an oscillatory
profile
\begin{equation}
I(t) = I_{0} \mathrm{exp} \left(-\alpha t\right) \sin \left( \beta t \right).
\label{eq:VillaCurrentEqn}
\end{equation}
Here, $I_0$ is the maximum current reached by the system, measured to
be $2.18\times 10^5$\,A, $\alpha$ is the damping factor and $\beta$ is
the damped frequency.  These are related to the properties of the
electrical circuit used to generate this current,
\begin{equation}
  \label{eq:villa-params}
  \alpha = \frac{R}{2L}, \quad \beta = \sqrt{\omega^2 - \alpha^2},
  \quad \omega = \sqrt{\frac {1}{LC}}
\end{equation}
where $\omega$ is the undamped frequency, $R$ is the resistance, $L$
the inductance and $C$ the capacitance.  These last three properties
are measured as $R = 24$\,m$\Omega$, $L = 2.9$\,$\mu$H and
$C = 26$\,$\mu$F respectively. 

For this validation test, the approach of VMB11 is used and a
pre-determined current density profile is provided,
\begin{equation}
  \label{eq:villa-curr-dens-rep}
  {\bf J} = -\frac{I\left(t\right)}{\pi r^{2}_{0}}e^{-\left(r /
      r_{0}\right)^{2}}\textbf{e}_{z}.
\end{equation}
Experimentally, the radius of the plasma arc was measured to be
between 1.5 and 2.5\,cm; a constant radius $r_0 = 2$\,cm is taken in
this work.  This allows the present implementation of the MHD
formulation to be validated in isolation.
\\

The set-up in Figure~\ref{fig:villa-expt} is modelled as a
two-dimensional cylindrical test, considering the substrate and the
electrode to be purely reflective boundaries for the material
properties of the plasma.  The current density profile given by
equation~(\ref{eq:villa-curr-dens-rep}) is applied between the
electrode and the substrate.

\begin{figure}[htbp]
  \centering
  \includegraphics[width=0.45\textwidth]{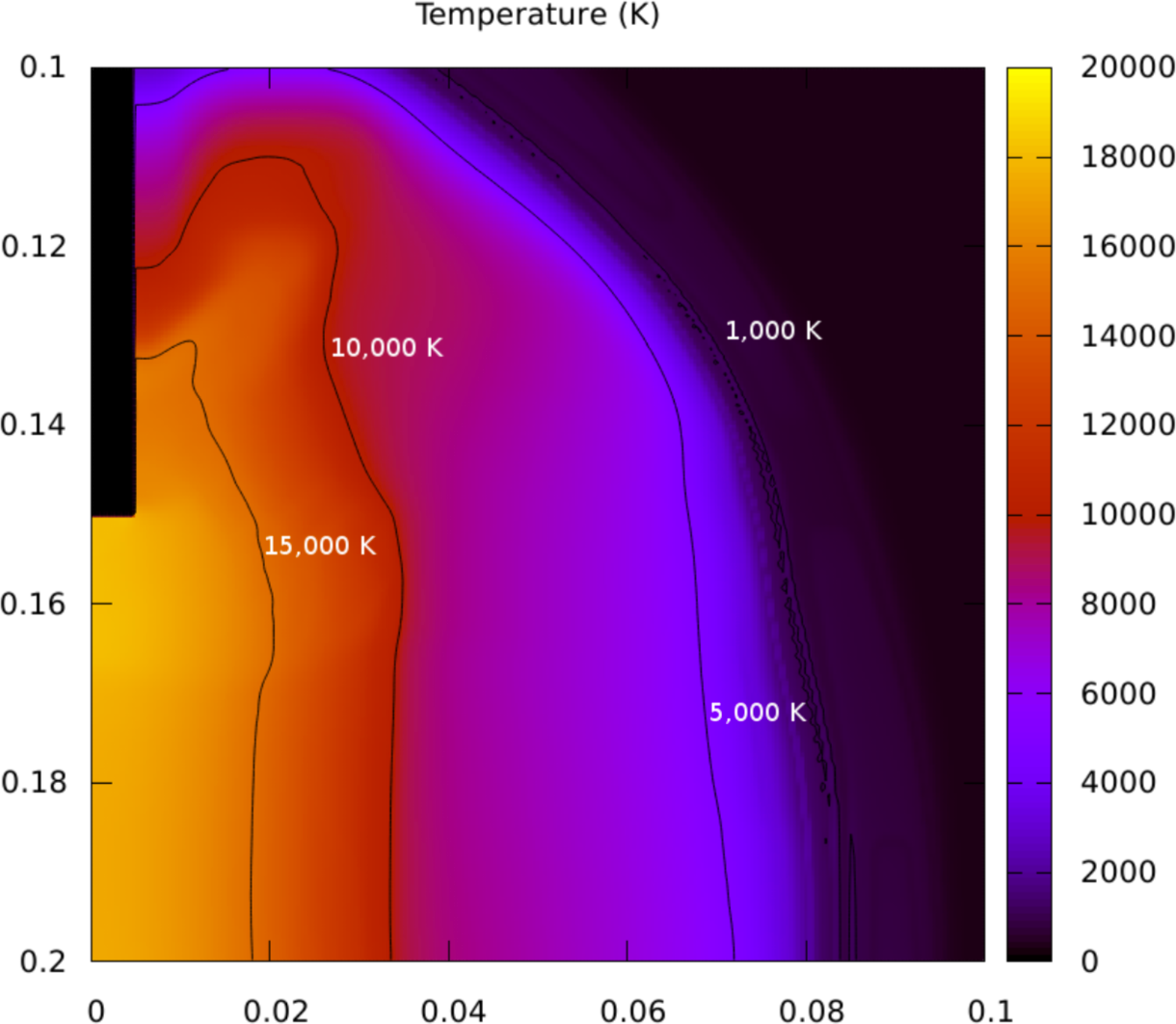}
  \caption{Comparison of the temperature profile after 48\,$\mu$s,
    comparable to the results of VMB11.  The results show
    qualitatively similar structures, though there are quantitative
    differences likely due to differences in the discrete-space
    implementation and uncertainty in the experimental parameters
    used. The outwards moving shock wave due to the initial formation
    of the arc is visible as tight contours up to around 5,000\,K.
    Behind this, there is a slower rise in temperature, and reflected
    features around the electrode are visible.}
  \label{fig:Villa3}
\end{figure}

In Figure~\ref{fig:Villa3}, the temperature profile is shown after
48\,$\mu$s, with the expansion of the arc, and the central
high-temperature region clearly visible.  The overall shape,
especially of the reflection around the electrode, is consistent with
the VMB11 results, though differences in the
temperatures profile are seen, which are likely to be due to
differences in the initialisation, and the numerical techniques used
(both for gridding and for evolution), between the models.

In order to model the pressure wave on the transducers, the technique
used in VMB11 is followed, which treats this as a separate problem to
the overall evolution of the plasma arc.  This avoids the need to
resolve the comparatively thin tubes within the simulation domain, and
preserves cylindrical symmetry.  The flow down these tubes is
simulated through a solution of the one-dimensional Euler equations,
modified to account for frictional effects of the tube edges.  These
equations are
\begin{align}
  \label{eq:euler-fric-1}
  \frac{\partial \rho}{\partial t} + \frac{\partial}{\partial x}
  \left(\rho v \right) & = 0 \\
  \label{eq:euler-fric-2}
  \frac{\partial}{\partial t} \left(\rho v\right) + \frac{\partial}{\partial x}
  \left(\rho v^2 + p \right) & = -\mathrm{sign}\left(\rho\right)
  \frac{1}{2}\frac{\lambda}{D} \rho v^2 \\
  \label{eq:euler-fric-3}
  \frac{\partial E}{\partial t} + \frac{\partial}{\partial x}
  \left[\left(E+p \right)v\right] & = \frac{\lambda}{D}\rho\abs{v^3}
\end{align}
where $\lambda = 0.018$ is the friction coefficient and $D= 1$\,cm is
the diameter of the tube.  The boundary conditions at the top of the
tube are specified by the properties of the plasma arc, and thus vary
with time.  These are given by
\begin{align}
  \rho = & \rho_p \nonumber \\
  \label{eq:fric-bc}
  v = & 0 \\
  p = &
  \begin{cases}
    p = p_p - \frac{1}{2}\kappa v, & \qquad v < 0 \\
    p = p_p, & \qquad v \ge 0
  \end{cases}\nonumber
\end{align}
where $\kappa = 0.43$ is the inlet pressure loss
coefficient~\cite{kotowski2011entrance}.  There is not substantial
flow of plasma into the tube, hence this system of equations can be
solved using a standard ideal gas, with $\gamma = 1.4$.

\begin{figure}[!ht]
  \centering
  \begin{subfigure}[tbp]{0.7\textwidth}
    \centering
    \includegraphics[width=\textwidth]{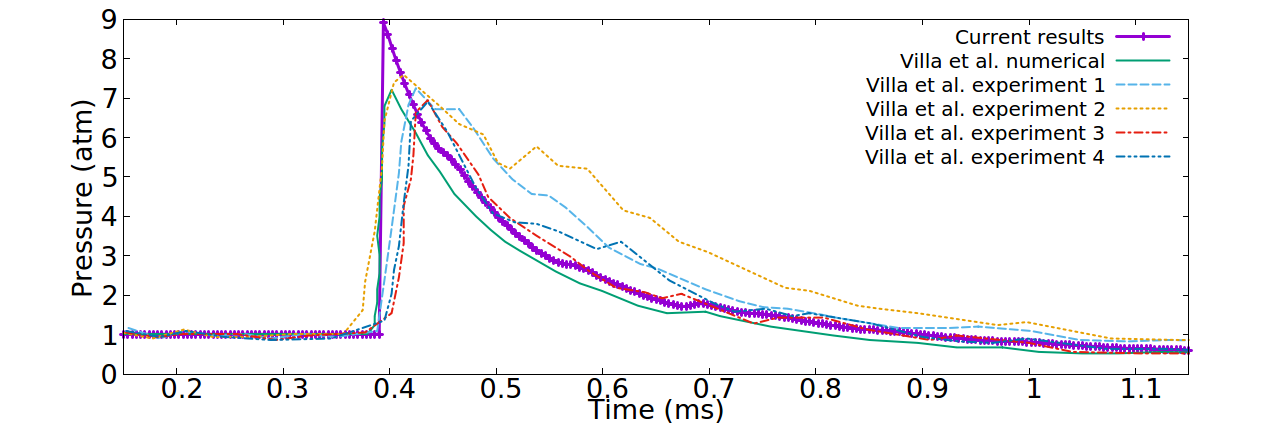}
    \caption{Gauge: 5\,cm}
    \label{fig:villa-gauge-a}
  \end{subfigure}

  \begin{subfigure}[tbp]{0.7\textwidth}
    \centering
    \includegraphics[width=\textwidth]{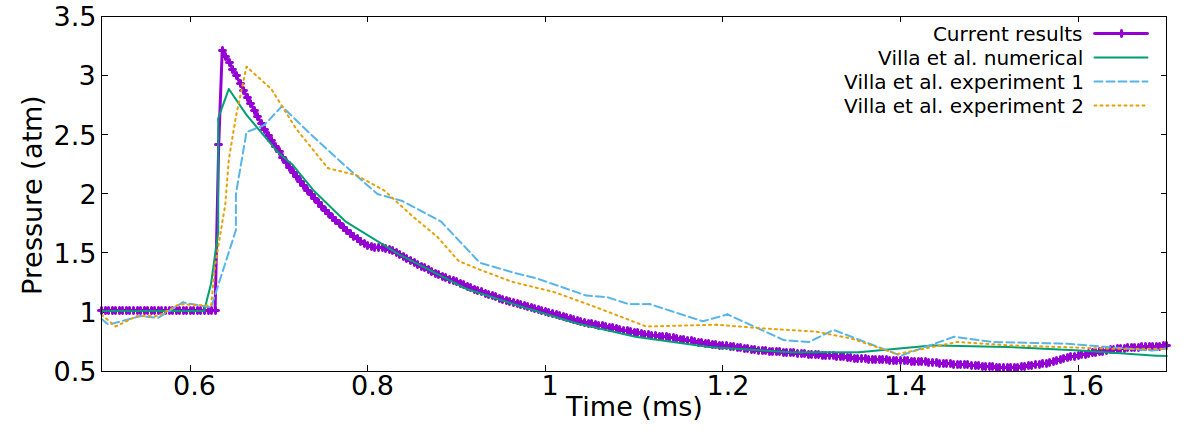}
    \caption{Gauge: 10\,cm}
    \label{fig:villa-gauge-b}
  \end{subfigure}

  \begin{subfigure}[tbp]{0.7\textwidth}
    \centering
    \includegraphics[width=\textwidth]{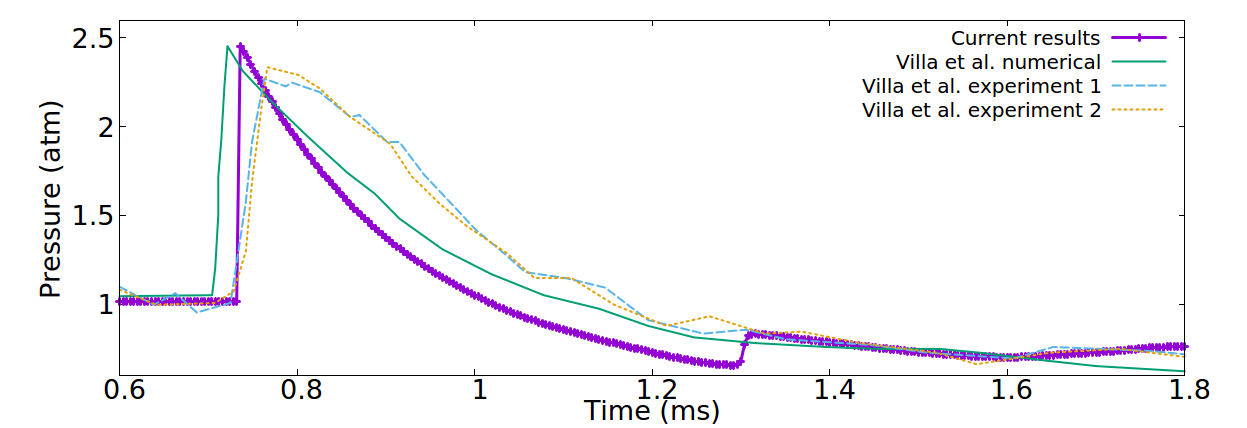}
    \caption{Gauge: 15\,cm}
    \label{fig:villa-gauge-c}
  \end{subfigure}

  \caption{Validation of the present plasma model against experimental
    and numerical results for flow travelling over a tubes 5\,cm,
    10\,cm and 15\,cm from the arc connection.  The purple curves are
    the present results, the green curves are the VMB11 numerical
    results and the remaining curves the VMB11 experimental results
    (between 2 and 4 for a given gauge).  The pressure peak and
    subsequent decay are reasonably well reproduced, given the
    uncertainty of the experimental conditions.  We note that the
    numerical models can over predict the pressure peak, since
    losses due to the assumption of a truly one-dimensional tube.}
  \label{fig:Villa4}
\end{figure}

In Figure~\ref{fig:Villa4}, the results are shown for the pressure
wave travelling down the three tube mounted 5, 10 and 15\,cm from the
arc attachment.  The present results are compared to both the VMB11
experimental and numerical results, with good comparison to both.  The
pressure peak is captured well, as is the rate of decay of the wave,
which levels out slightly below atmospheric pressure.  The simulation
results typically give a slight overestimation of the pressure, due to
the one-dimensional assumption used to model the tubes.  As a result,
any dissipation of the wave, due to interaction with the tube walls is
lost.

The results shown in this section validate the present numerical
implementation of the equations governing plasma dynamics.  In
particular, they show that the correct behaviour of the outwards
moving shock wave is correctly captured.

\subsection{Validation of the fully coupled system}
\label{sec:valid-full-coupl}

When an elastoplastic substrate is incorporated, assumptions as to the
shape of the current density profile can no longer be made.  In this
case, equation~(\ref{eq:cons-j}) is solved for current density across
the entire domain.  When solving the complete coupled system, the
interaction with the substrate affects the evolution of the plasma
arc.  In order to validate the present model in this case, the results
are compared to the experimental data of M16.  Through high-speed
imaging of the arc attachment and early evolution up to around
40\,$\mu$s, both the width of the plasma arc, and the progression of
the shock wave generated by the arc formation, could be measured.

In order to validate the present implementation, two substrate
configurations are considered, aluminium and an isotropic
approximation to a carbon composite material (hereafter referred to as
the isotropic composite).  The electrical conductivity of these two
materials differs substantially, aluminium has
$\sigma = 3.2 \times 10^{7}$\,S/m whilst the carbon composite the
isotropic material is based on has
i$\sigma = 1.6 \times 10^{4}$\,S/m~\cite{tholin2015numerical}.
Experimental results show that for a material with lower conductivity
material area is larger.  This is due to the electrical conductivity
of this material being comparable to that of the plasma, hence the
extended arc connection offers a less resistive path for the current
flow.

The initial data for this model uses the set-up described in
Figure~\ref{fig:arc-substrate-example}, which incorporates the
electrode for current input within the domain.  The initial data for
this problem uses a pre-heated arc region at the centre of the domain
of 8,000\,K with a 2\,mm diameter.  For these tests, either a 1\,mm
(for comparison of an aluminium substrate to M16) or 2\,mm (elsewhere)
thick substrate is considered, located 40mm below the electrode.  A
direct current application through the electrode following a
D-component waveform is used, as defined in the document
ARP~5412B~\cite{ARP5412B}.  A polynomial fit is used to the current
profile recorded by M16.

\begin{figure}[!ht]
  \centering
  \includegraphics[width=1.0\textwidth]{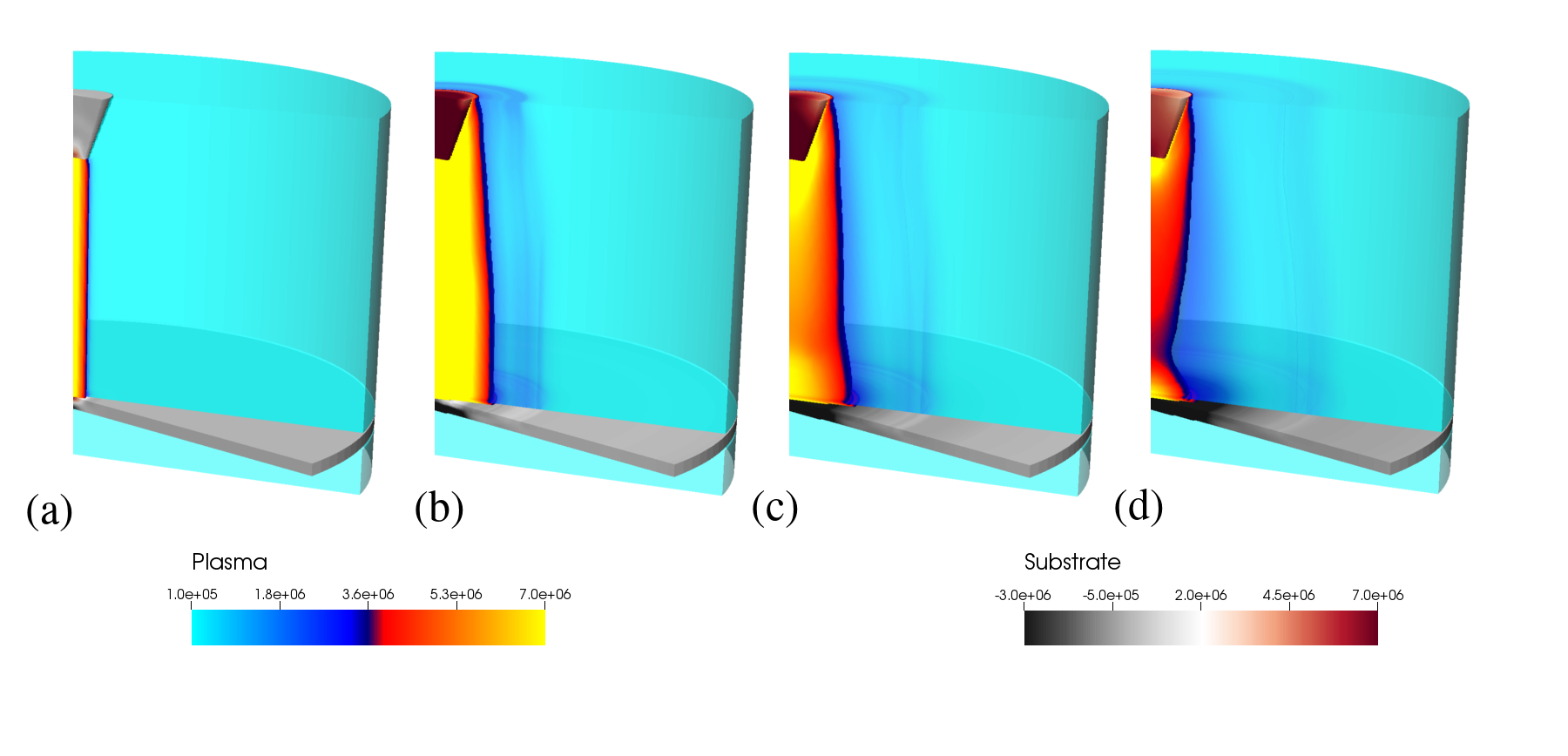}
  \caption{Pressure evolution for an arc attachment to an aluminium
    substrate at times (a) 1$\mu$s, (b) 10$\mu$s, (c) 15$\mu$s and
    (d) 20$\mu$s.}
  \label{fig:Test2AluPressure}
\end{figure}

Figure~\ref{fig:Test2AluPressure} shows the evolution of the pressure
profile in the arc and a 2\,mm thick aluminium substrate.  After
1$\mu$s, Figure~\ref{fig:Test2AluPressure}~(a), a pressure loading is
evident on the aluminium substrate. Although the current input is
still reasonably low at this early time, there is sufficient energy
input such that the pressure at the centre of the arc has doubled from
the initial pre-heated value. As the arc evolves over time,
Figure~\ref{fig:Test2AluPressure}~(b-d), the highest pressure remains
in the centre of the arc, with the majority of the pressure loading on
the substrate occurring here.  Away from the centre of the arc, the
higher pressure associated with the initial shock wave moving radially
away from the centre of the arc is also visible as a darker blue
region in the plasma.  A corresponding wave moves through the
substrate as the shock wave imparts a loading effect, though this is
substantially lower than at the centre of the arc.  Whilst the
pressure within the plasma arc must strictly stay positive, it is
noted that within the substrate, negative values are experienced.
This is because pressure is a component of the stress tensor, and a
solid material can sustain tension, as a result of rarefaction waves.
When considering overall damage effects, the magnitude of the stress
within the substrate is an important criterion.

The effect of the dynamic current density profile is clearly visible;
the arc in Figure~\ref{fig:Test2AluPressure} does not maintain a
cylindrical shape.  Current density gradient in the $z$-direction
leads to a higher pressure directly beneath the arc, but also at the
attachment point, where high pressures, due to reflected material, are
seen.

\begin{figure}[!ht]
  \centering
  \includegraphics[width=1.0\textwidth]{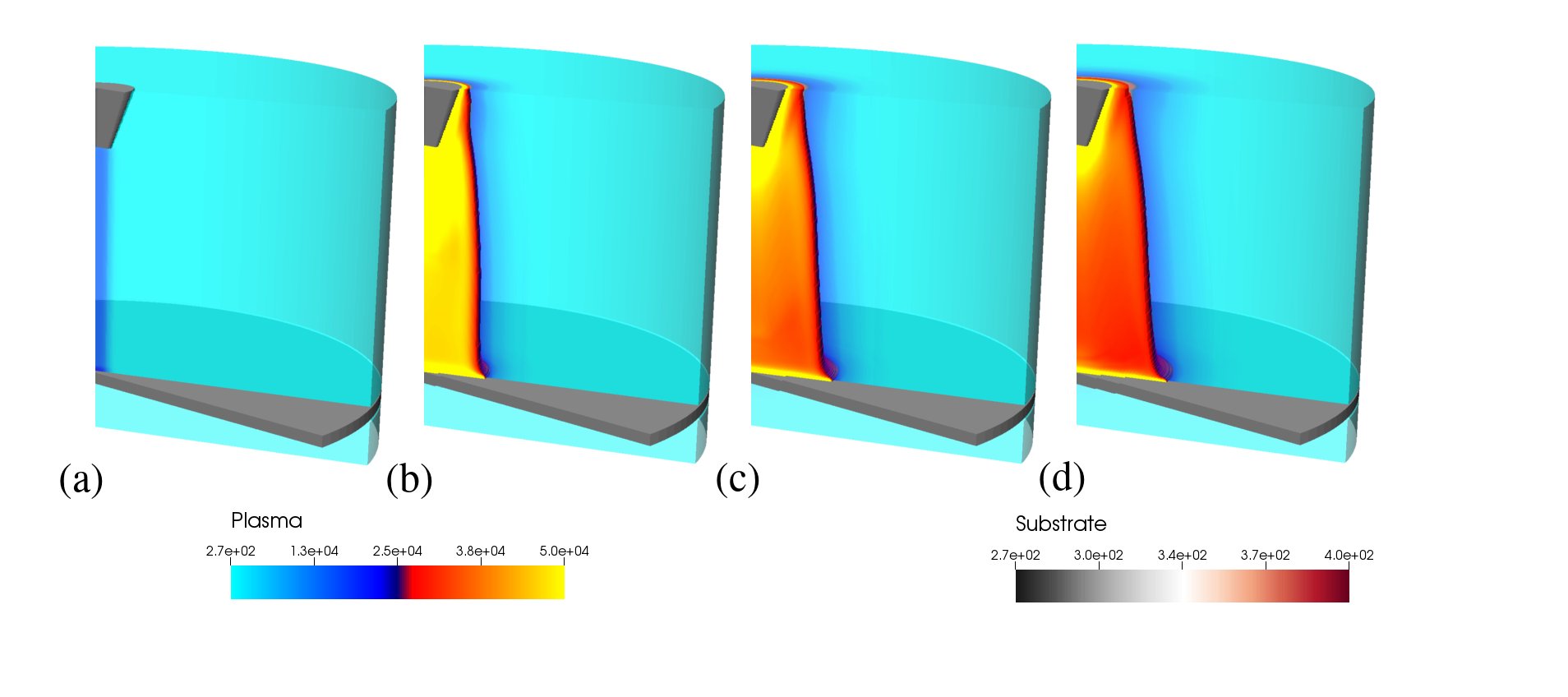}
  \caption{Temperature evolution in both the plasma arc and the
    aluminium substrate at times of (a) 1$\mu$s, (b) 10$\mu$s, (c)
    15$\mu$s and (d) 20$\mu$s.}
  \label{fig:Test2AluTemperature}
\end{figure}

The temperature in the plasma and aluminium substrate as the arc
evolves is shown in Figure~\ref{fig:Test2AluTemperature}~(a-d), at
corresponding times to the pressure images in
Figure~\ref{fig:Test2AluPressure}.  The hottest temperature regions
remain in the centre of the domain, where conductivity, and thus Joule
heating, is greatest.  The temperature is also plotted in the
substrate, however, due to the high conductivity of the aluminium
substrate in this test, the energy density deposited in the substrate
is comparatively low at this timescale.  The overall rise in
temperature over the timescales considered is less than 1\,K.  Over
longer timescales ($\mathcal{O} (1)$\,s), temperature rise would be
governed by diffusive and conductive behaviour, in addition to longer
term Joule heating effects from the long continuous current, and this
leads to the minor damage resulting from lightning strike.

\begin{figure}[!ht]
  \centering
  \includegraphics[width=0.4\textwidth]{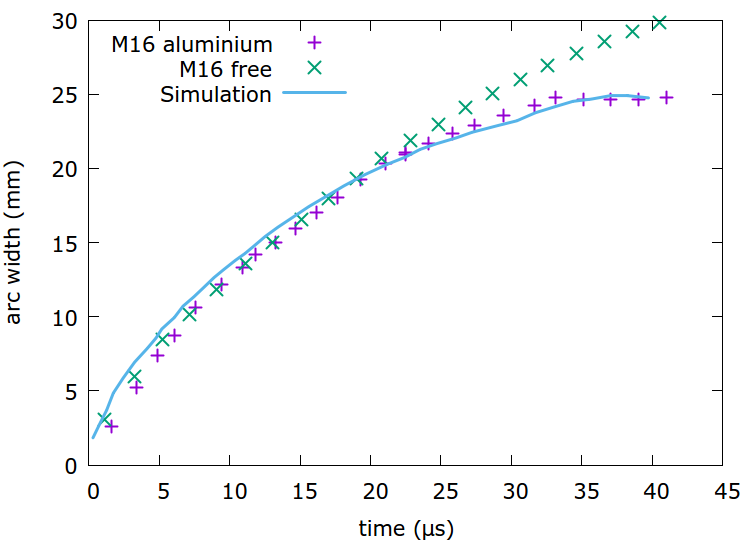}
  \includegraphics[width=0.4\textwidth]{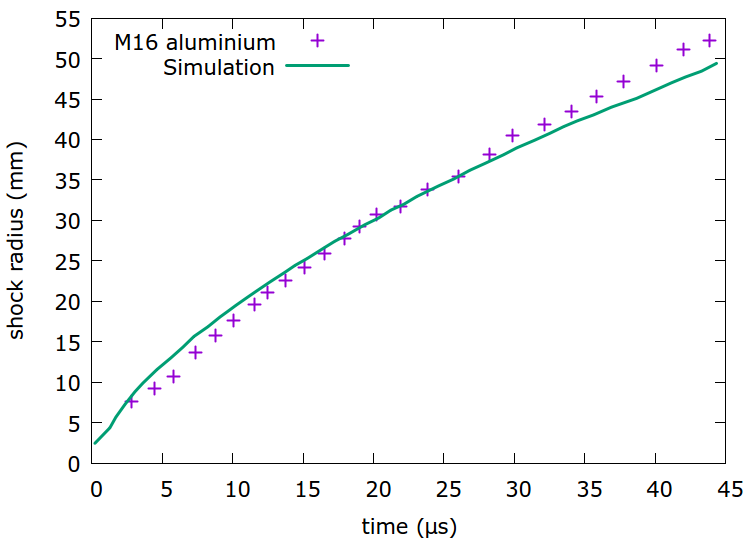}

  \caption{The left plot shows the comparison of the measurements of
    the arc width for experiment ($+$) and simulation (solid line).
    The numerical results are shown to correctly capture the evolution
    of the arc on an aluminium substrate.  For comparison, the
    experimentally measured width of an arc between two electrodes is
    plotted ($\times$), showing the substrate has a clear effect on
    the evolution itself.  The right plot shows the comparison of the
    experimentally measured ($+$) and computational (solid line) shock
    wave propagation.  It is clear that the present model captures
    this behaviour well.}
  \label{fig:martins-arc-comp}
\end{figure}

These results qualitatively compare well with the images obtained in
M16.  By taking measurements of the arc width and the shock
progression, a quantitative comparison to the experimental results can
be made.  In the left half of Figure~\ref{fig:martins-arc-comp}, the
present numerical results are compared to M16 for arc attachment to a
1\,mm thick sheet of aluminium.  The numerical results match the
experimentally measured widths, demonstrating the interaction with the
substrate can be correctly captured.  This is further evident since
the width of an arc between two electrodes with no substrate present
(sometimes referred to as a free arc) is plotted.  There is
significant difference in the evolution of the widths of the free arc
compared to when a substrate is present after around 20\,$\mu$s, which
is correctly captured by the present model.

The M16 experiment was also able to capture the evolution of the shock
wave generated by the plasma arc through optical changes in a
patterned background.  In the right half of Figure~\ref{fig:martins-arc-comp} the
present numerical results for the propagation of the shock wave are
compared, and again we find good agreement between the present model
and experimental studies.
\\

\begin{figure}[!ht]
  \centering
  \includegraphics[width=1.0\textwidth]{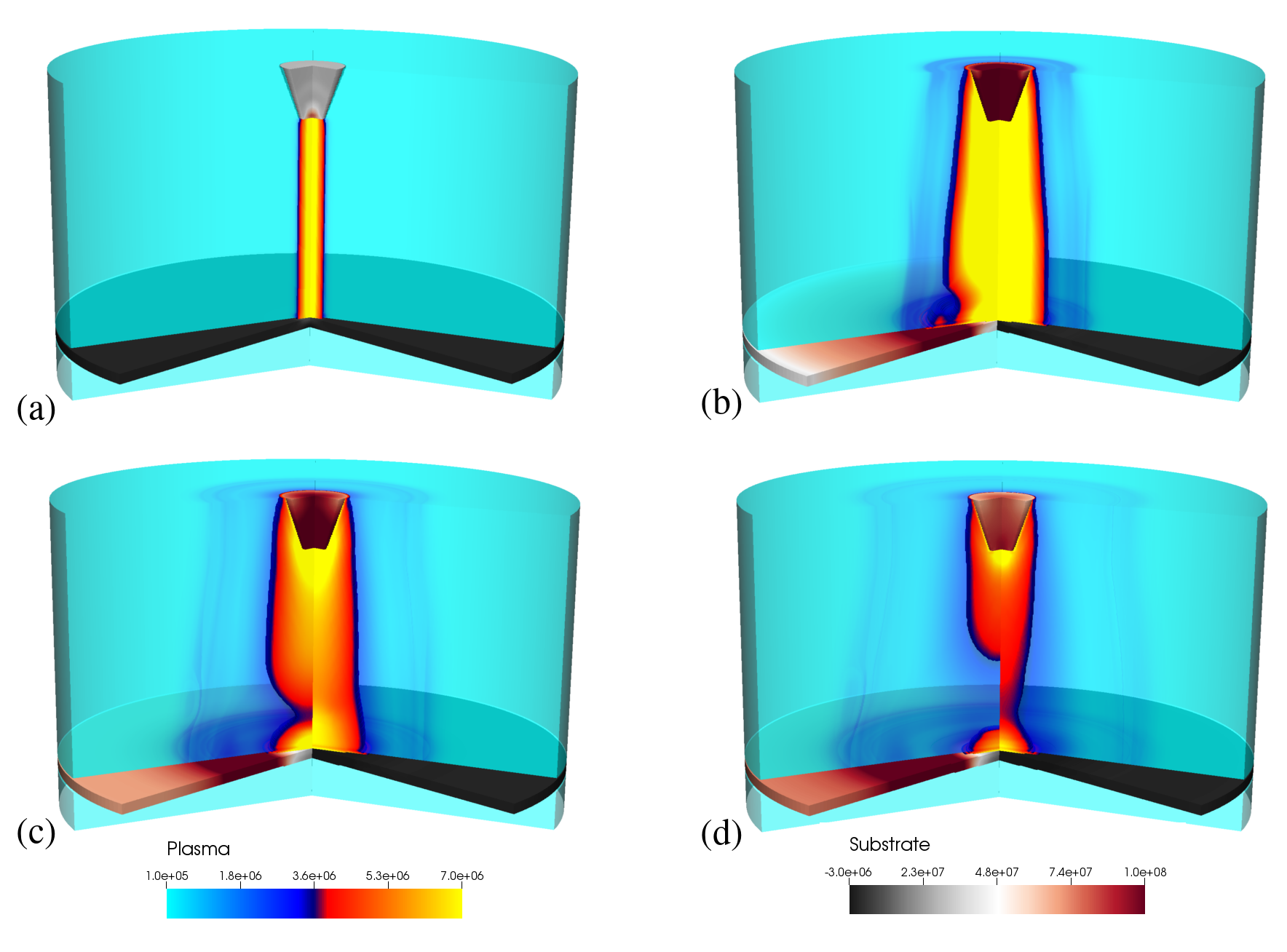}
  \caption{Pressure evolution of plasma arc attachment to the
    isotropic composite substrate (left) and to the aluminium
    substrate (right) at times of (a) 1 $\mu$s, (b) 10 $\mu$s, (c) 15
    $\mu$s and (d) 20 $\mu$s.}
  \label{fig:Test2CarbonAluPressure}
\end{figure}

Arc attachment to the isotropic composite substrate is now considered
and results are compared directly to the results shown for an
aluminium substrate.  The full modelling of a carbon composite
material requires an anisotropic description of the alignment of the
fibres comprising the substrate to be incorporated within the equation
of state.  This is currently beyond the capabilities of the present
model, though by making an isotropic approximation to a composite
material, the effect the substrate conductivity has on the arc can be
considered.  This isotropic model is approximately equivalent to a
description of the composite material in the direction of the fibres.

The comparison between a low-conductivity substrate and aluminium are
presented with both results shown on the same plot, the isotropic
composite substrate plotted at the left of the central axis, and the
aluminium substrate on the right.  In order to visualise the
differences between the simulation, the same plot ranges are always
used for both materials.

In Figure~\ref{fig:Test2CarbonAluPressure} the evolution of the
pressure for attachment to the isotropic composite substrate is shown.
There are clear differences between the evolution profiles, both
within the arc and the substrate, compared to attachment to aluminium.
Within the arc, the behaviour local to the electrode is largely
unchanged, it is clear that the differences originate due to the
interaction with the substrate.  In
Figure~\ref{fig:Test2CarbonAluPressure}, there is a high-pressure
region close to the surface of the substrate, which then has `pinch'
type behaviour directly above it.  This serves to exacerbate the
gradient in pressure down the arc, visible in
Figures~\ref{fig:Test2CarbonAluPressure}~(c-d).  Additionally, in
Figure~\ref{fig:Test2CarbonAluPressure}~(c), it is clear that the
change in behaviour at the substrate surface leads to a faster
shock-propagation speed, though at the top of the domain, the shock
speed remains similar to the case of an aluminium substrate.  Within
the substrate, the pressure loading is substantially higher.  This is
a result of greater energy deposition in the substrate through the
Joule effect in equation~(\ref{eq:solid-form-2}).  This is further
evidenced by the location of the low pressure region in the substrate
beneath the arc.  This region is substantially larger than that
beneath the arc attachment to aluminium, and in fact, exists even
where pressure loading is highest.  This suggests that there are
additional effects contributing to the pressure increase within the
substrate, subsequent plots show that the high pressure region is
correlated to high current density, and hence Joule heating.

\begin{figure}[!ht]
  \centering
  \includegraphics[width=1.0\textwidth]{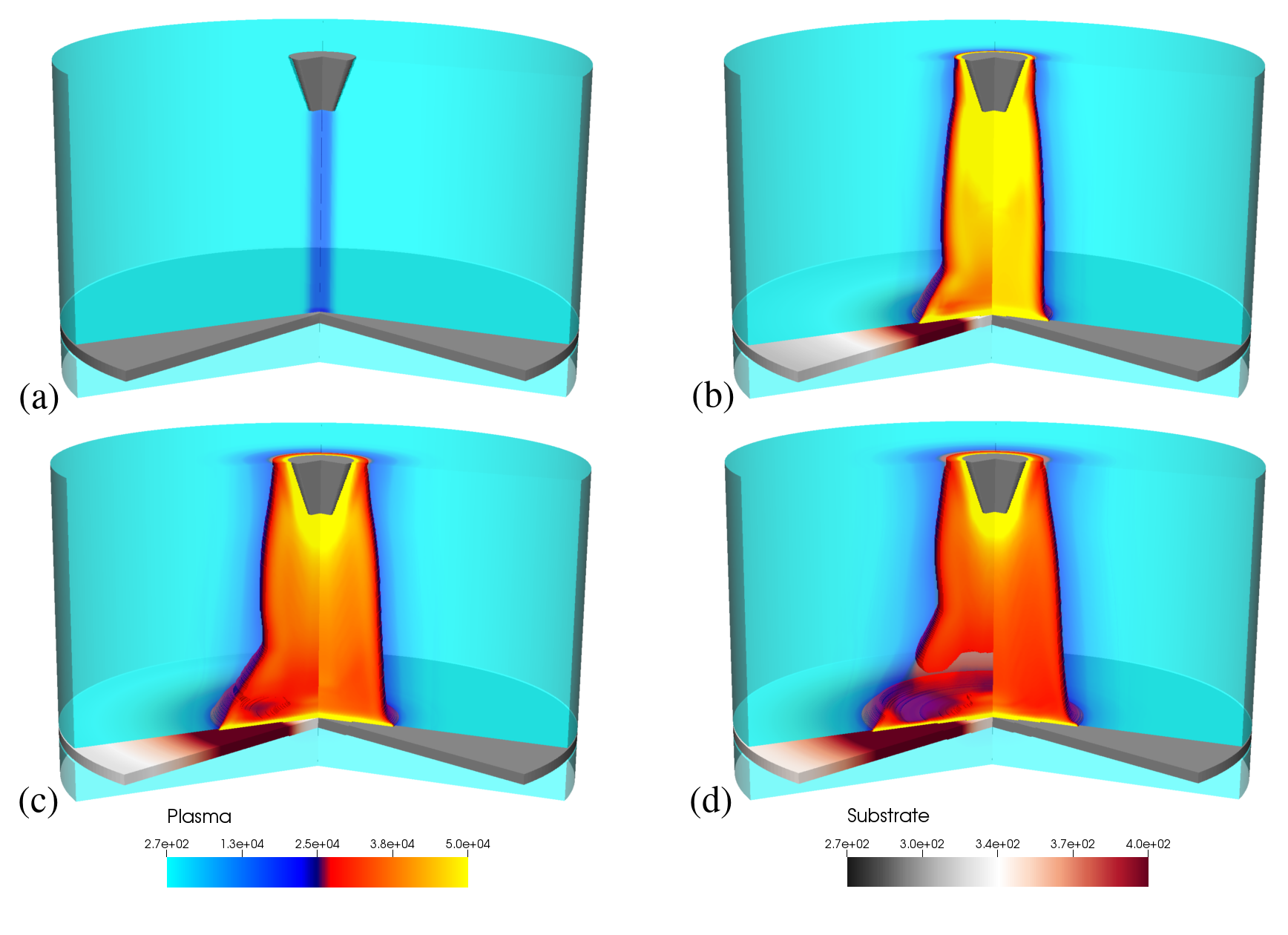}
  \caption{Temperature evolution of the plasma arc with an isotropic
    composite substrate (left) and an aluminium substrate (right) at
    times of (a) 1 $\mu$s, (b) 10 $\mu$s, (c) 15 $\mu$s and (d) 20
    $\mu$s.}
  \label{fig:Test2CarbonAluTemperature}
\end{figure}
The differences in the pressure loading of the aluminium and isotropic
composite substrates are shown in the temperature field over the same
time period in Figure~\ref{fig:Test2CarbonAluTemperature}.  As the
plasma arc develops, the radius of the plasma arc close to the top
surface of the substrate is greater than that of the aluminium.  Where
the arc radius is large, a lower temperature region is seen,
particularly on the outer edges of the arc.  Such a region is visible
in the optical emission results of Tholin {\em et al}.\ suggesting the
correct coupling is captured between the isotropic composite substrate
and the arc.  The optical emission is closely coupled to conductivity,
which is itself dependent on temperature, thus a comparison can be
made between these two results.

The temperature within the substrate is also plotted, and it is clear
that there is a noticeable increase for the isotropic composite
material.  Due to the energy deposition through the Joule heating
effect, there is a corresponding rise in temperature.  It is clear
that there is in an increase associated with the leading edge of the
arc.
\\

\begin{figure}[!ht]
  \centering
  \includegraphics[width=0.4\textwidth]{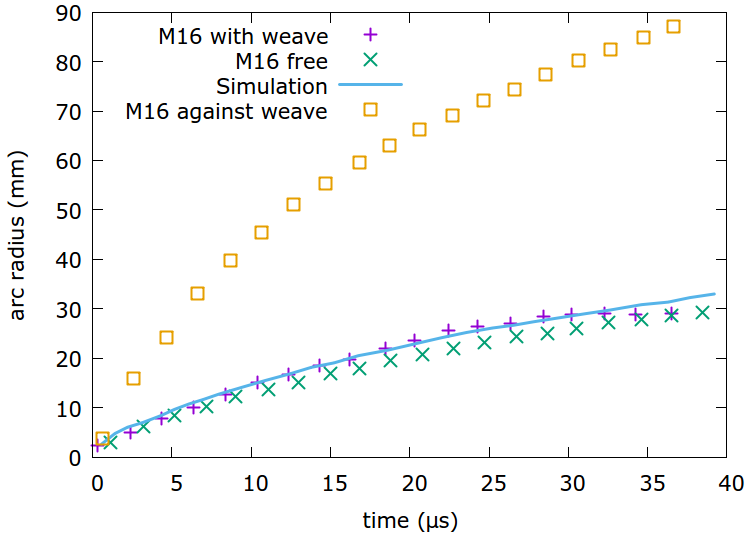}
  \includegraphics[width=0.4\textwidth]{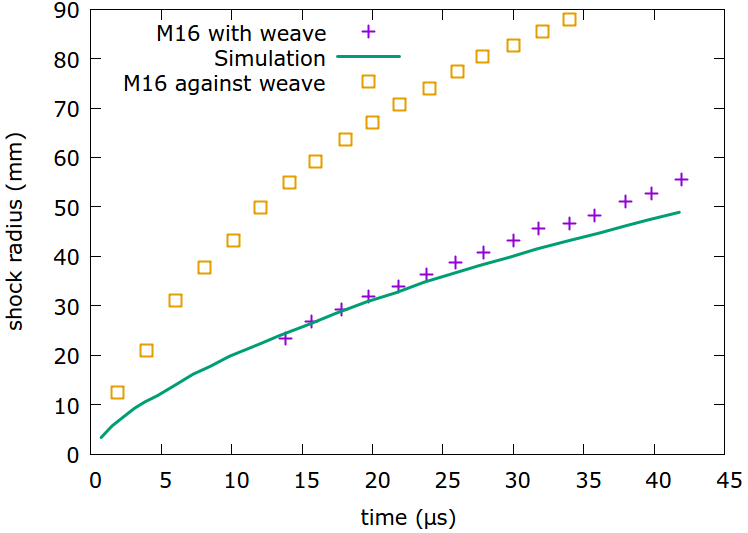}
  \caption{The left plot shows the comparison of the measurements of
    the arc width for experimental attachment to carbon composite ($+$
    and $\Box$) and simulation (solid line) of attachment to a
    low-conductivity substrate.  The experimental model shows both arc
    widths along the carbon weave direction ($+$) and perpendicular to
    the weave ($Box$).  The present numerical results correctly
    capture the evolution of the arc on along the weave direction.
    For comparison, the experimentally measured width of an arc
    between two electrodes is plotted ($\times$).  The right plot
    shows the comparison of the experimentally measured ($+$ and
    $\Box$) and computational (solid line) shock wave propagation
    against a low-conductivity substrate.  Again, the two
    experimental results correspond to measurements along the carbon
    weave ($+$), and perpendicular to it ($\Box$).  It is clear that
    the present model captures the shock expansion corresponding to
    the direction along the weave well.}
  \label{fig:martins-carbon-width}
\end{figure}

As with the arc attachment to aluminium, the M16 experimental results
can be used to further validate this model.  In
Figure~\ref{fig:martins-carbon-width} the arc width obtained from both
experiment and simulation is compared.  There are two experimental
values for arc width plotted, one in the direction aligned to the
carbon weave, and one perpendicular to it.  It is clear that the
isotropic approximation captures the behaviour aligned to the weave
well (this is the preferential direction for current to travel).
Again the arc width of the free arc is plotted, and it is now clear
that there is significant differences in the arc width for this
low-conductivity case and the aluminium attachment shown in
Figure~\ref{fig:martins-arc-comp}.

In the right half of Figure~\ref{fig:martins-carbon-width} the
expansion of the shock wave for the experiment and simulation is
compared.  As for the arc width, two experimental values are obtained,
depending on the orientation of the recording equipment to the carbon
weave.  As before, the current isotropic model of a low-conductivity
substrate is found to correspond well to the behaviour in the
direction of the carbon weave.

\section{Multi-layered substrates}
\label{sec:multi-layer-substr}

\begin{figure}[!ht]
  \centering
  \includegraphics[width=1.0\textwidth]{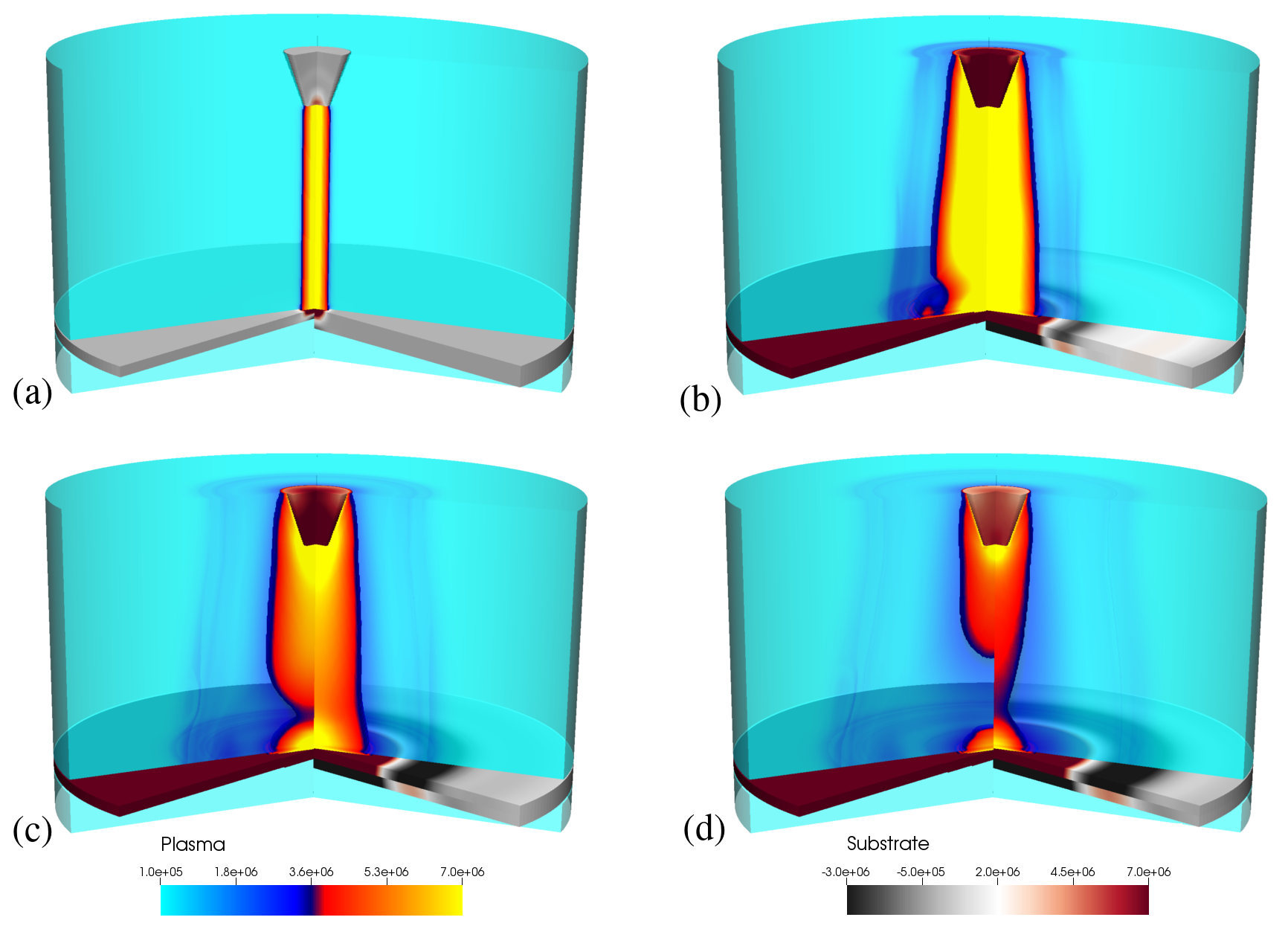}
  \caption{Pressure comparison for arc interaction with an isotropic
    composite substrate in isolation (left) and a dual layered
    substrate (right) at times of (a) 1 $\mu$s, (b) 10 $\mu$s, (c) 15
    $\mu$s and (d) 20 $\mu$s.  The same pressure range is chosen for
    both sets of results to enable direct comparison.}
  \label{fig:Test3CarbonAluPressure}
\end{figure}

The multimaterial nature of the present model allows for the plasma
arc to interact with materials that are not necessarily on the surface
of the substrate.  In this section, a test case which investigates the
effects of layering materials with different electrical conductivities
is considered.  This is constructed such that an isotropic composite
substrate is placed on top of a sheet of aluminium.  By placing the
high-conductivity substrate as the bottom layer in this scenario, the
effects of an embedded layer used within current carbon composite
materials is considered; it is expected that this layer can form a
preferential path for the current flow.

Each layer has a thickness of 2\,mm, giving a total substrate
thickness of 4\,mm.  To initialise the plasma, a pre-heated region
directly connecting the electrode to the substrate is again included.
In order to ascertain the effects of the dual layered substrate, the
results are compared to that of a single-layer isotropic composite
substrate.  Therefore, if the results are governed only by the top
material, it would be expected that there is no difference visible
between these two cases.

Figure~\ref{fig:Test3CarbonAluPressure} shows the pressure evolution
within the plasma and the substrate, compared to the situation where
there is just isotropic composite present.  The pressure profile in
both plasma and substrate are clearly effected by the presence of the
dual layering. The plasma arc does not show the `pinch' feature, and
subsequent expansion close to the surface, when the aluminium layer is
included.  The overall shape follows that of the single aluminium
substrate shown in Figure~\ref{fig:Test2AluPressure}.  It is clear
that this change in behaviour is also true for the expansion of the
shock wave above the dual layered substrate.  Additionally, the high
pressure loading on the dual substrate is now confined to the area
directly beneath the electrode.  There is still a higher pressure
within the isotropic composite substrate than for a single aluminium
sheet, but the extent is confined to only a small region.

\begin{figure}[!ht]
  \centering
  \includegraphics[width=1.0\textwidth]{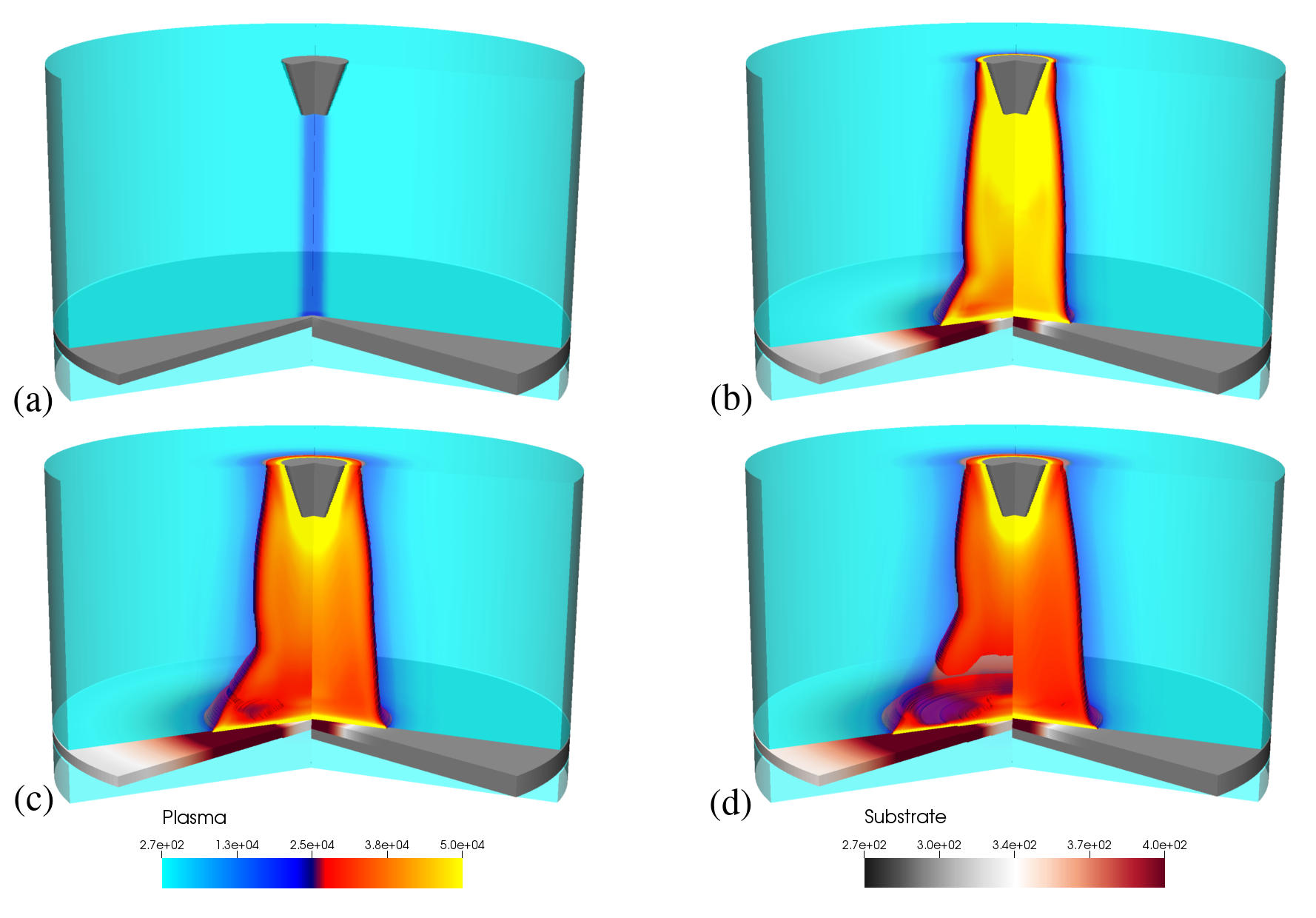}
  \caption{Temperature evolution for arc attachment to an isotropic
    composite substrate (left) and a dual layered substrate (right) at
    times of (a) 1 $\mu$s, (b) 10 $\mu$s, (c) 15 $\mu$s and (d) 20
    $\mu$s.}
  \label{fig:Test3CarbonAluTemperature}
\end{figure}
The reduction in the radial extent of the high pressure region as a
result of adding an aluminium layer shown in
Figure~\ref{fig:Test3CarbonAluPressure} may be expected to yield a
reduction in substrate heating.
Figure~\ref{fig:Test3CarbonAluTemperature} shows the temperature
evolution corresponding to this pressure behaviour. As expected, the
temperature profile follows the general trend of the pressure profile
with the high temperature region being significantly reduced in radial
extent through the introduction of the aluminium layer. Once again,
the shape of the temperature profile is comparable to the single
aluminium layer, shown in Figure~\ref{fig:Test2AluTemperature}.
As the pressure profile suggests, there is a high temperature in the
dual layered substrate restricted to the initial arc attachment point.
Away from this, there is no substantial increase in the temperature of
the substrate, a clear contrast to the single isotropic composite
substrate case.  The reduction in temperature in the substrate is a
consequence of less energy being deposited in this layer due to Joule
heating, which suggests that the current flow is predominantly through
the aluminium layer to the ground site, instead of through the low
conductivity layer.

\begin{figure}[!ht]
  \centering
  \includegraphics[width=0.6\textwidth]{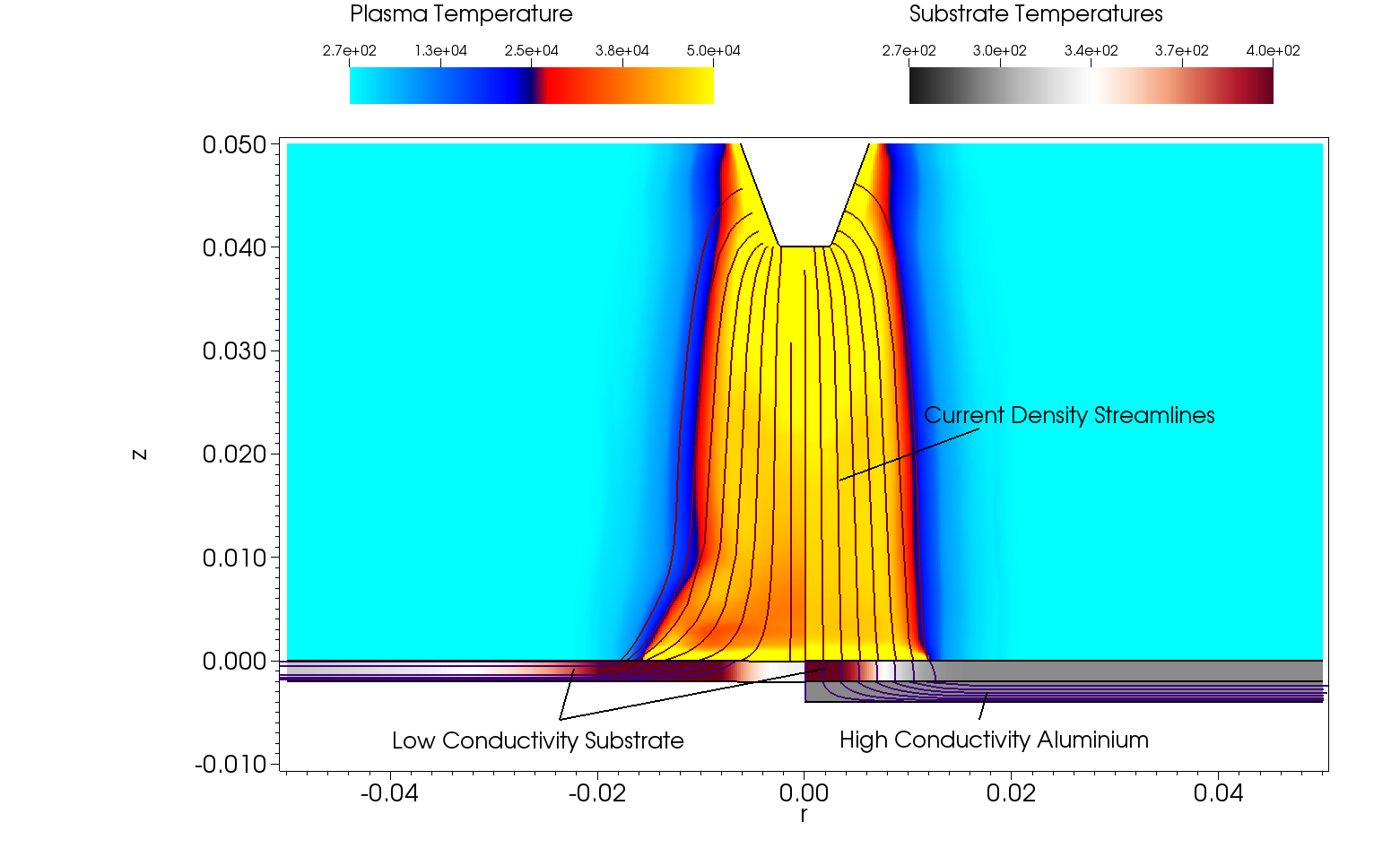}
  \caption{Annotated slice through the temperature field at t=10\,$\mu$s
    comparing an isotropic composite substrate (left) to a dual layered
    substrate (right). Current streamlines are overlaid to show the path
    of current from the electrode to the grounding location.}
  \label{fig:Test3AnnotatedTemperature}
\end{figure}

In Figure~\ref{fig:Test3AnnotatedTemperature} the current density
streamlines at $t = 10$\,$\mu$s are shown, i.e.\ at the same time as
Figure~\ref{fig:Test3CarbonAluTemperature}~(b).  These are plotted
over temperature field, to show the path of the current, and how other
variables are follow this behaviour closely.  For the isotropic
composite substrate, there is a clear radial component to the
streamlines within the plasma as they approach the substrate.  As
expected, in this case, they then attach directly to the ground site.
When the aluminium layer is included, the difference in path for the
streamlines is clear.  They now flow much more directly into the
aluminium substrate.  It is at this point they turn radially towards
the ground site, and remain within the aluminium layer.  This explains
the similarities between the results for a single aluminium substrate
in Figures~\ref{fig:Test2AluPressure} and
\ref{fig:Test2AluTemperature} and the dual layered results presented
in this section.

The results in this section demonstrate that the model presented is
capable of accurately coupling the interaction of a plasma arc with a complex
arrangement of substrate layers.  The complete properties of these
layers will govern the behaviour of the arc attachment, and not just
the properties of the top layer.  

\section{Temperature-dependent conductivity}
\label{sec:temp-depend-cond}

The fully coupled nature of the present model means that behaviour
within the substrate can alter the shape of the arc.  This is most
obvious when the energy input into the substrate alters the electrical
conductivity of the material.  For high-conductivity substrates such
as aluminium, these effects are negligible over the timescales
considered, since there is very little change in the temperature of
the substrate.  However, the significant energy deposition into a
low-conductivity substrate is sufficient to alter the material
properties, such as electrical conductivity, over a short timescale.
It would then be expected that this will alter the interaction of the
arc with the substrate, demonstrating a true two-way non-linear
coupling.

A test case is considered in which a temperature-dependent electrical
conductivity for a carbon composite material is taken from Guo {\em et
  al}.~\cite{8260383}, and applied to the isotropic composite used in
this work.  The conductivity of the material decreases close to
linearly with increased temperature, hence a line is fit through the
experimental data according to
\begin{equation}
  \label{eq:t-dep-cond}
  \sigma_0 +  \alpha T
\end{equation}
where $\alpha = -17.115$\,S\,m$^{-1}$\,K$^{-1}$.  The reference
conductivity for this material is
$\sigma_0 = 1.45\times 10^4$\,S\,m$^{-1}$, and thus two test cases are
considered; firstly where the conductivity within the substrate is
constant at the value $\sigma_0$, and secondly where it obeys
equation~(\ref{eq:t-dep-cond}).

\begin{figure}[!ht]
  \centering
  \includegraphics[width=0.45\textwidth]{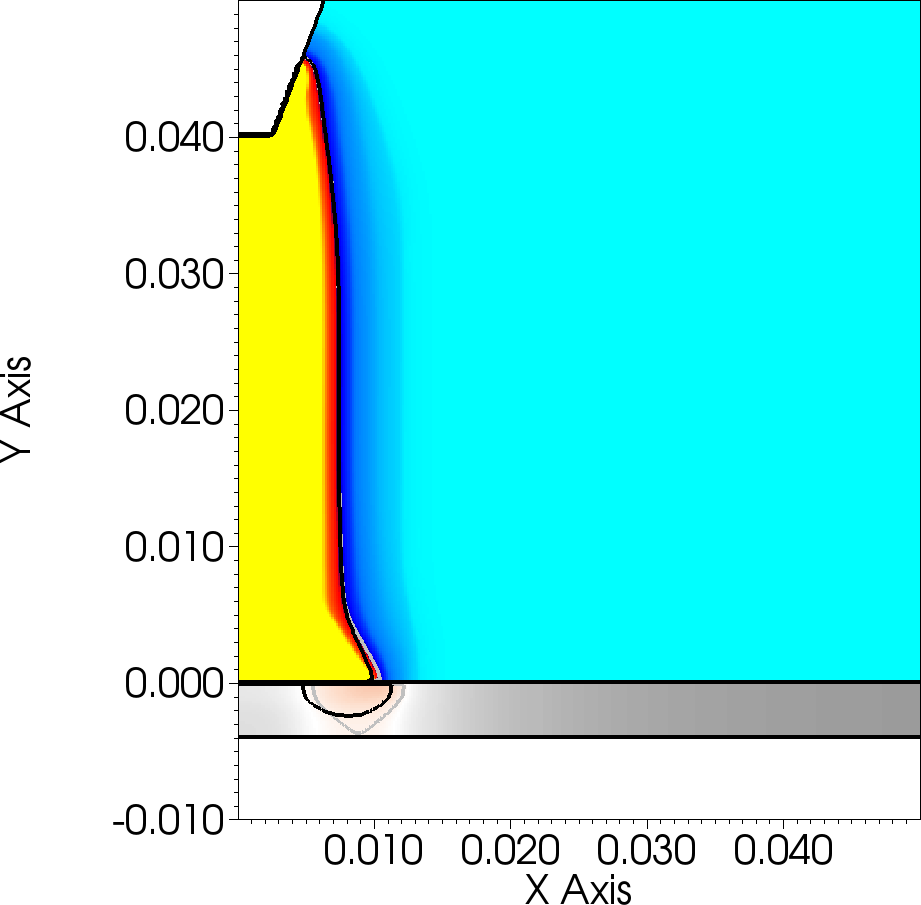}
  \includegraphics[width=0.45\textwidth]{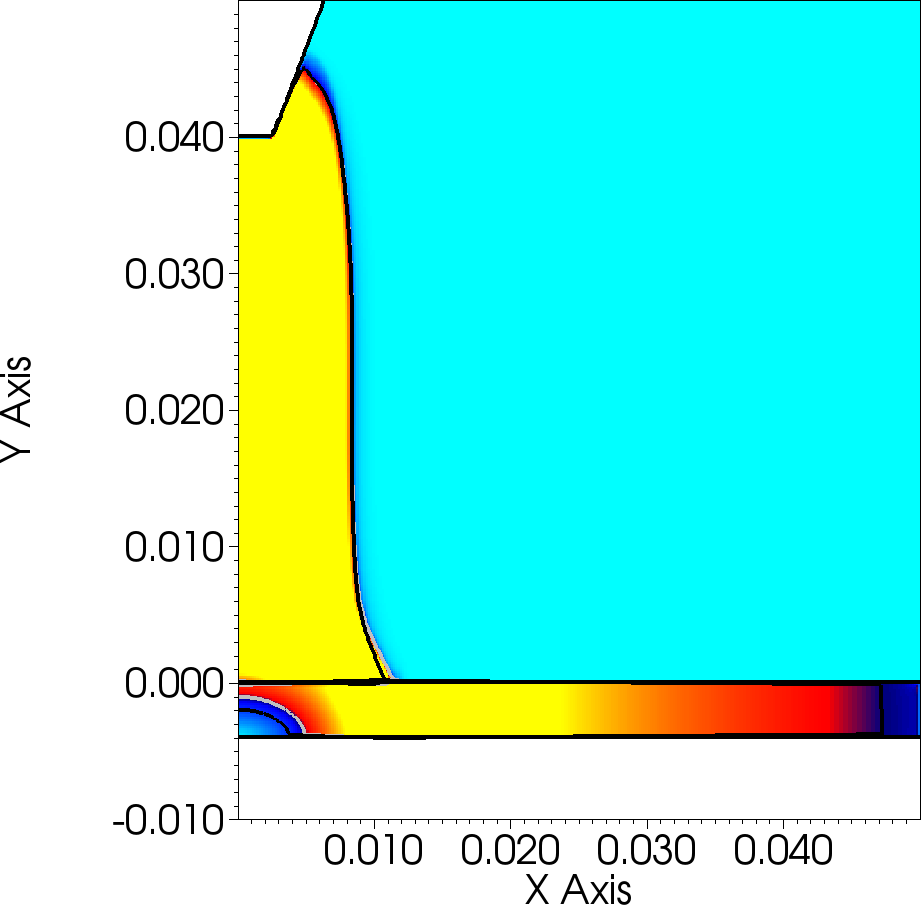}

  \includegraphics[width=0.8\textwidth]{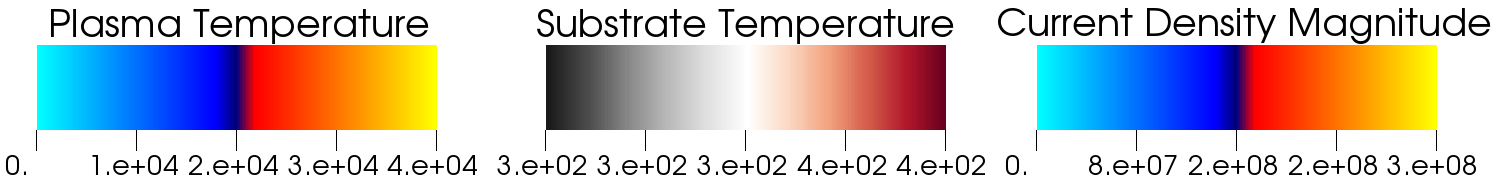}
  \caption{Temperature (left) and current density magnitude (right)
    images for a substrate with a temperature dependent conductivity
    after 5.4\,$\mu$s.  Constant value contours are shown in grey;
    $2\times10^4$\, K for temperature in the plasma arc, 340\,K for
    the substrate and $1.5\times 10^8$\,A/m$^{-2}$ for current density
    magnitude everywhere.  The black contours show the corresponding
    values for a constant-conductivity substrate.  At this time, the
    arc profiles are comparable, though slightly wider at the
    attachment point with a temperature-dependent conductivity.
    However, it is clear that the substrate is heating more rapidly,
    and the extent of this is further radially outwards.}
  \label{fig:var-cond-plots}
\end{figure}
\begin{figure}[!ht]
  \centering
  \includegraphics[width=0.45\textwidth]{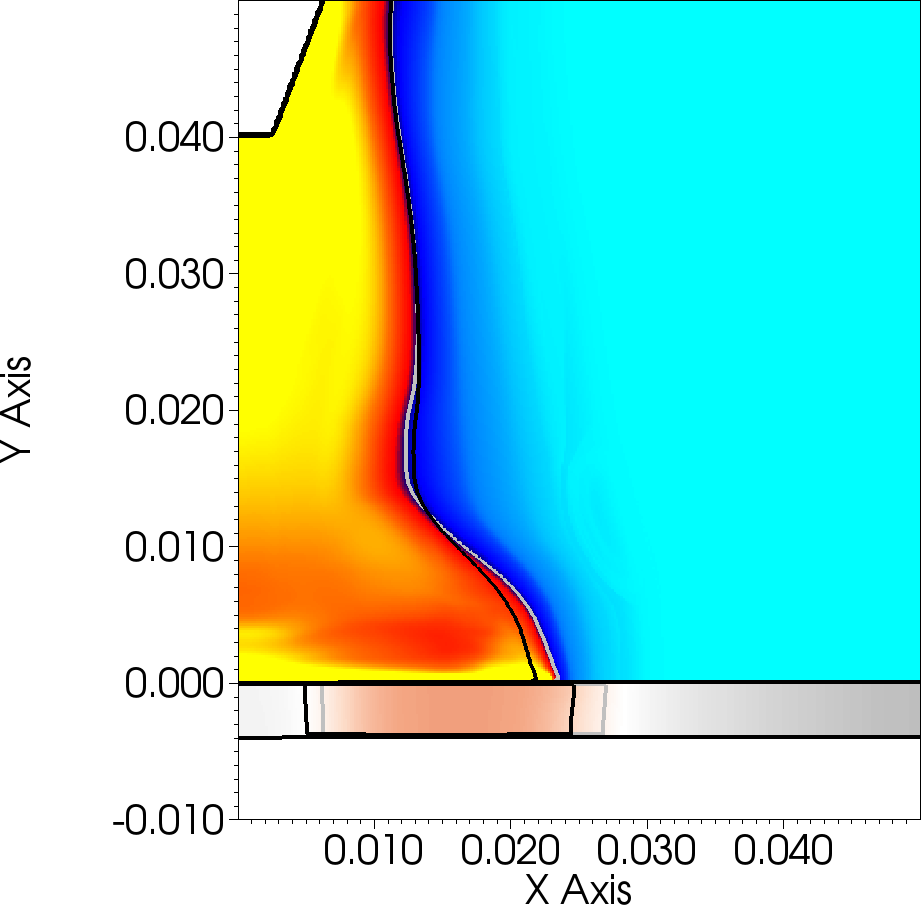}
  \includegraphics[width=0.45\textwidth]{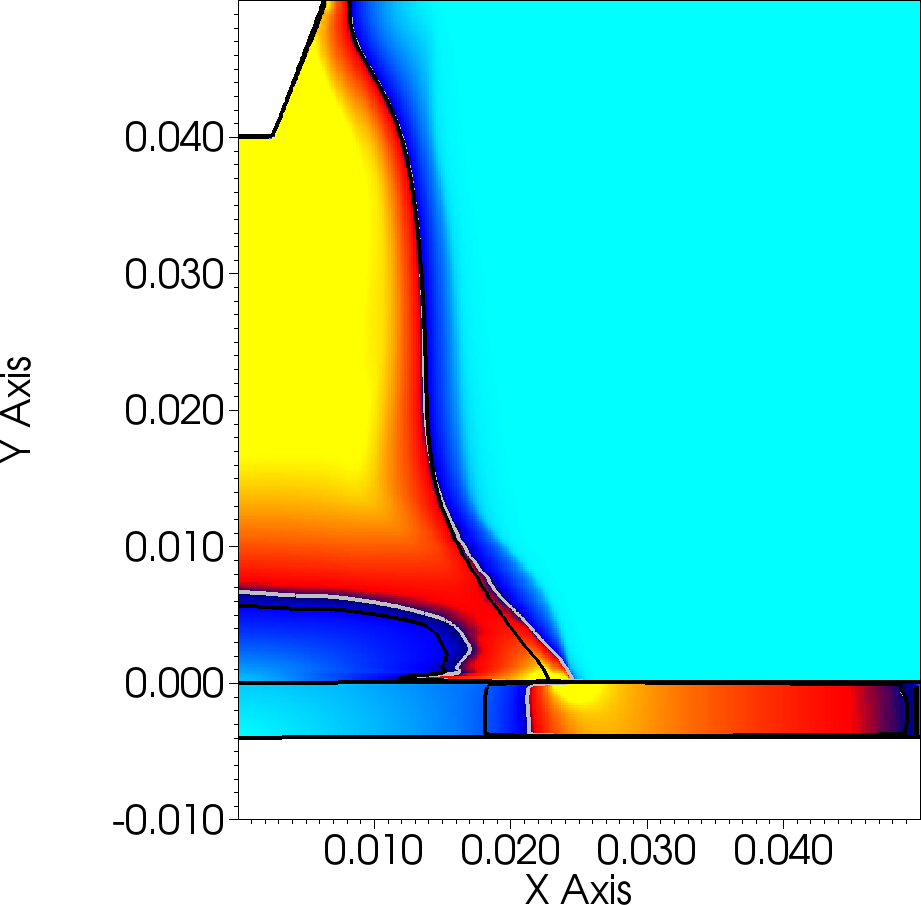}

  \includegraphics[width=0.8\textwidth]{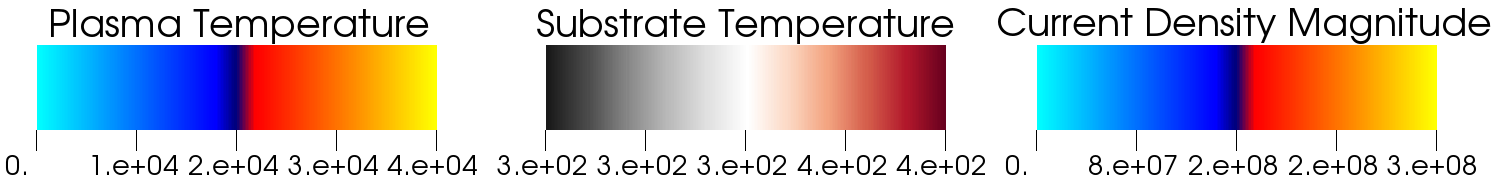}
  \caption{Temperature (left) and current density magnitude (right)
    images for a substrate with a temperature dependent conductivity
    after 15\,$\mu$s.  Constant value contours are shown in grey;
    $2\times10^4$\, K for temperature in the plasma arc, 340\,K for
    the substrate and $1.5\times 10^8$\,A/m$^{-2}$ for current density
    magnitude everywhere.  The black contours show the corresponding
    values for a constant-conductivity substrate.  The effects of
    the temperature-dependent substrate are now clear; the arc is wider
    directly above the substrate, and the path of current flow is also
    wider as a result.  This leads to a larger heated
    region in the substrate, and the current flow enters at a greater
    radial distance.}
  \label{fig:var-cond-plots-2}
\end{figure}

The effects of the temperature-dependent conductivity are shown after
5.4\,$\mu$s in Figure~\ref{fig:var-cond-plots}, and after 15\,$\mu$s
in Figure~\ref{fig:var-cond-plots-2}.  These plots show the effect of
the temperature-dependent substrate properties, both on the
temperature and current density magnitude profiles.  For each figure,
two contours are plotted, one for the constant conductivity model, and
one for the temperature dependent conductivity.  Both contours are
plotted at the same value; $2\times10^4$\, K for temperature in the
plasma arc, 340\,K for the substrate and $1.5\times 10^8$\,A/m$^{-2}$
for the current density in both materials.

In Figure~\ref{fig:var-cond-plots}, the current input to the system is
close to its peak value.  At this stage, there is little difference in
the arc profile, its evolution is being governed primarily by this
current input.  It is, however, slightly wider where it attaches to
the substrate.  The effects within the substrate itself are more
noticeable, both in temperature and current density.  The contour at
340\,K is deeper in the case of a varying substrate conductivity, and
the primary path of the current through the substrate is clearly
radially further outwards.  In Figure~\ref{fig:var-cond-plots-2}, the
difference between the two cases are now clear in the plasma arc, as
well as the substrate.  The increase in width of the arc attachment is
more pronounced, and this leads to a decrease in width, seen in the
temperature profile midway between the substrate and the electrode.
The extended path of the current through the arc, causing this greater
attachment area, is visible in the current density magnitude.  Within
the substrate, the heated region is both wider, and radially further
out.  Additionally, the maximum temperature in this region is higher
in the case of variable conductivity.  The current density profile in
the substrate again shows a greater radial distance of the attachment.

These results demonstrate successful simulation of the feedback
between the plasma arc and the substrate.  The Joule heating effect
imparting energy into the substrate alters its properties, and hence
the optimal path for current to take.  As a result, the shape of the
plasma arc is changed, in this case moving further outwards.  In this
particular case, including the temperature dependent properties of the
substrate could show that greater damage occurs to the substrate than
would otherwise be predicted, due to the larger area of effect, and
the greater temperatures reached within the substrate.

\section{Conclusions}
\label{sec:conclusions}

In this work a methodology is presented for the simulation of the
non-linear interaction between a plasma arc and an elastoplastic
substrate.  Within this model, each material (plasma, substrate(s),
electrode and air) is modelled with its own set of PDEs, all of which
are written in the same hyperbolic form.  These are solved
simultaneously, rather than by means of a coupled, `co-simulation'
approach.  By using level set methods and the ghost fluid method,
dynamic boundary conditions are provided between each material through
the solution of mixed-material Riemann problems.  This allows the
entire system to be simulated on a single computational grid, and
capture the non-linear interactions between the materials present.  To
the best of the authors' knowledge, this is the first time the ability
to capture the dynamic feedback of evolution within the substrate
affecting the plasma arc has been demonstrated.

This approach is validated through comparison to experimental studies,
where it is demonstrated that the present model can capture the
correct growth of a plasma arc dependent on the composition of the
substrate.  The electrical conductivity of the substrate should alter
the width of a plasma arc, with a lower conductivity leading to a
wider attachment area; it was demonstrated that this behaviour could
be replicated on aluminium, and on a low-conductivity isotropic
approximation to a carbon composite.  Additionally it was possible to
validate the location of the shock wave generated by the arc attachment
against experimental studies.

The capability of the model to deal with a substrate comprising two
distinct materials was then evaluated, which is representative of the
layered approached used in modern aircraft design.  A trial case was
considered with a layer of aluminium mounted beneath a layer of the
isotropic composite substrate.  In this case, it was shown that the
present model could capture the current flow through the low
conductivity material into the aluminium, and thus restrict the growth
of the plasma arc.  In practical applications, the present model would
be able to capture the properties of a high-conductivity mesh,
diverting current flow from a substrate primarily composed of a
low-conductivity material such as CFRP.

In order to demonstrate the full non-linear behaviour of the present
model, a test case where the evolution of the substrate had a feedback
effect on the plasma arc was considered.  A temperature-varying
electrical conductivity was taken from experimental work on CFRP.  In
this case, as the substrate temperature increased, the electrical
conductivity decreased.  This study indicated that lowering the
substrate conductivity caused the plasma arc to expand, local to the
substrate, as would be expected for the optimal current path.
Additionally, the change in electrical conductivity of the substrate
lead to greater heating through the Joule effect.  This two-way
dynamic feedback between the arc and the substrate can be used to
better identify where damage to a substrate configuration may occur.

Future development using this multi-physics framework will focus on
the development of the equation of state used within the plasma arc
and air, and on techniques to improve the radiative source term within
the arc in a computationally efficient manner.  The substrate model
shall also be developed further, considering the effects of a
dielectric layers such as paint, which inhibit the growth of the
plasma arc, and additionally the conductive behaviour of the substrate
will be altered to better model CFRP.

\section*{ Acknowledgements} 

The authors acknowledge the funding support of Boeing Research \&
Technology (BR\&T) through project number SSOW-BRT-L0516-0569.  We
would also like to thank Micah Goldade of BR\&T for technical input
throughout the work, and Carmen Guerra-Garcia (BR\&T and currently at
the Massachusetts Institute of Technology) for suggestions on how best
to model behaviours of a plasma.

\bibliographystyle{elsarticle-num} 

\bibliography{ms.bib}

\end{document}